\documentclass[pra,twocolumn]{revtex4-1}
\usepackage{hyperref} 

\usepackage{graphicx} 
\usepackage{subfig}
\usepackage{color}
\usepackage{filecontents}
\usepackage{soul}
\usepackage{mathrsfs}
\usepackage{tikz,bm} 
\usepackage{circuitikz}
\usepackage{tikz}
\usepackage{physics}
\usepackage{mathtools}
\usepackage[T1]{fontenc}
\usepackage[outline]{contour}
\usepackage{relsize}

\usetikzlibrary{shapes.geometric}
\usetikzlibrary{decorations.pathmorphing}
\usetikzlibrary{shapes.arrows, fadings}
\usetikzlibrary{arrows,snakes,shapes}

\definecolor{mycolor}{rgb}{1,0.2,0.3}
\definecolor{brightgreen}{rgb}{1.0, 1.0, 1.0}
\definecolor{britishracinggreen}{rgb}{0.0, 0.26, 0.15}
\definecolor{cadmiumgreen}{rgb}{0.0, 0.42, 0.24}
\definecolor{ceruleanblue}{rgb}{0.16, 0.32, 0.75}
\definecolor{darkelectricblue}{rgb}{0.33, 0.41, 0.47}
\definecolor{darkpowderblue}{rgb}{0.0, 0.2, 0.6}
\definecolor{dt}{rgb}{1.0, 0.66, 0.07} %darktangerine
\definecolor{emerald}{rgb}{0.31, 0.78, 0.47}
\definecolor{palatinatepurple}{rgb}{0.41, 0.16, 0.38}
\definecolor{pastelviolet}{rgb}{0.8, 0.6, 0.79}
\definecolor{br}{rgb}{0.5, 0.05, 0.01}
\definecolor{chosen_color}{RGB}{3, 207, 252}

\pgfdeclareverticalshading{pgf@lib@fade@north}{100bp}
{color(0bp)=(pgftransparent!0);
 color(30bp)=(pgftransparent!0);
 color(50bp)=(pgftransparent!80); 
 color(65bp)=(pgftransparent!100);
 color(100bp)=(pgftransparent!100)}%
\pgfdeclarefading{exponentialtop}{%
\pgfuseshading{pgf@lib@fade@north}%
}

\pgfdeclareverticalshading{pgf@lib@fade@north}{100bp}
{color(0bp)=(pgftransparent!100);
 color(35bp)=(pgftransparent!100);
 color(50bp)=(pgftransparent!80); 
 color(70bp)=(pgftransparent!00);
 color(100bp)=(pgftransparent!00)}%
\pgfdeclarefading{exponentialbottom}{%
  \pgfuseshading{pgf@lib@fade@north}%
}

\newcommand{\be}{\begin{equation}}
\newcommand{\ee}{\end{equation}}
\newcommand{\bea}{\begin{eqnarray}}
\newcommand{\eea}{\end{eqnarray}}
\newcommand*{\myeqref}[2][Eq.~]{%
\hyperref[{#2}]{#1(\ref*{#2})}%
}
\def\equationautorefname#1#2\null{%
Eq.#1(#2\null)%
}

\usepackage{dcolumn}
\usepackage{bm}
\usepackage{amssymb,amsmath}
\usepackage{color}
\usepackage{float}
\usepackage{tikz}
\usetikzlibrary{arrows}

\definecolor{DarkGreen}{rgb}{0,0.6,0.2}

\usepackage{mathrsfs}
\begin{document}
\title{Tripartite Entanglement in Multimode Cavity Quantum Electrodynamics}
\author{Nishan Amgain$^{1}$} 
\author{Mahir Rahman$^{1,2}$}  
\author{Umar Arshad$^{1,3}$}
\author{Fernando Romero$^{1,4}$} 
\author{Emil Sayahi$^{1}$}
\author{Imran M. Mirza$^{1}$}
\email{mirzaim@miamioh.edu}
\affiliation{$^{1}$Macklin Quantum Information Sciences,Department of Physics, Miami University, Oxford, OH 45056, USA\\
$^{2}$Department of Computer Science, Purdue University, West Lafayette, IN 47907, USA\\
$^{3}$Department of Physics and Astronomy, Rice University, Houston, TX 77005, USA\\
$^{4}$Department Materials Science and Engineering, Ohio State University, Columbus, OH 43210, USA}
\date{\today}

%%===================================================%%
%%                   Article Abstract                %%
%%===================================================%%
\begin{abstract}
We numerically investigate the generation and dynamics of tripartite entanglement among qubits (quantum emitters or atoms) in multimode cavity quantum electrodynamics (cQED). Our cQED architecture features three initially unentangled excited two-level quantum emitters confined within a triangle-shaped multimode optical cavity, which later become entangled due to a Jaynes-Cummings-like interaction. Using the tripartite negativity measure of entanglement and fidelity with respect to the genuine tripartite entangled state (Greenberger-Horne-Zeilinger (or GHZ) state, to be precise), we analyze the impact of the number of cavity modes, qubit locations, and losses (spontaneous emission from qubits and photon leakage from the cavity mirrors) on the generated entanglement. Our key results include the presence of two kinds of retardation effects: one resulting from the time it takes for photons to propagate from one qubit to another, and the other to complete one round trip in the cavity. We observed these retardation effects only in multimode cavities, with the exciting possibility of controlling the collapse and revival patterns of tripartite entanglement by altering the qubit locations in the cavity. Furthermore, the impact of losses on the generated entanglement and the dependence of maximum entanglement on the total number of modes yield results that surpass those reported for single and two excitations. With recent advances in circuit quantum electrodynamics, these findings hold promise for the development of entanglement-based quantum networking protocols and quantum memories. 
\end{abstract}

%%%%%%%%%%%%%%%%%%%%%%%%%%%%%%%%%%%%%%%%%%%%%%%%%%%%%%%

\maketitle

%%===================================================%%
%%         Sec. I: Article Introduction              %%
%%===================================================%%
\section{\label{sec:I} Introduction}  
Since the seminal EPR paper (Einstein, Podolsky, and Rosen) of 1935 \cite{einstein1935can}, quantum entanglement has been a central topic of debate concerning the foundations of quantum mechanics. With the landmark work of John Bell in the 1960s \cite{bell1964einstein}, followed by experimental tests of Bell's inequalities by Clauser \cite{freedman1972experimental}, Aspect \cite{aspect1982experimental}, and Zeilinger \cite{kafatos2013bell} in the 1970s and 1980s, entanglement went beyond being merely a philosophical topic and was realized as an experimental fact. With the dawn of the second quantum revolution in the early 1990s \cite{dowling2003quantum}, entanglement was recognized as a valuable resource for information in various types of quantum technologies. These include quantum cryptography \cite{bennett1992experimental}, where entanglement is used to secure communication channels, and quantum teleportation \cite{bouwmeester1997experimental}, a process that allows the transfer of quantum information from one location to another. Since then, entanglement has remained a central topic in a wide range of protocols and applications in quantum computation and quantum information processing \cite{bennett2000quantum}. 

Since no quantum system exists in isolation, when entangled particles interact with their environment, entanglement is known to show extremely fragile behavior \cite{isar2009entanglement}. In many such scenarios, the temporal decay of entanglement occurs in a very short time compared to what is required for any real-world quantum technological application, which poses severe challenges in the use of entanglement as an information resource \cite{yu2009sudden, carvalho2004decoherence, aolita2008scaling}. In this context, cQED platforms — high-quality factor optical cavities coupled with two-level atoms or quantum emitters \cite{shore1993jaynes} — provide an excellent setup for studying strong light-matter interactions. Such interactions are the key to the generation and survival of qubit-qubit entanglement, which can be theoretically analyzed using the Jaynes-Cummings (JC) and JC-like models \cite{oliveira2008protecting}. 

In the standard paradigm of JC models, cavities are assumed to support a single optical mode \cite{larson2021jaynes}. However, manufacturing optical cavities capable of supporting exactly one mode can be a challenging task experimentally. Additionally, recent developments in quantum computing based on circuit quantum electrodynamics have revealed novel quantum effects when quantum emitters are simultaneously coupled to multiple optical modes (see, for example, Ref.~\cite{naik2017random, chakram2021seamless}). Motivated by these considerations, this paper focuses on the problem of qubit-qubit entanglement generation and dynamics in the multimode version of JC-like models. A survey of the literature indicates that the problem of multimode JC models at the single and two-photon levels has been studied in the past \cite{ficek2012effect, bougouffa2013effect, gulfam2013creation}. In this work, by going beyond the problems of bipartite qubit-qubit entanglement induced by single or two photons, we study the tripartite entanglement of three qubits in multimode JC-like models while working in the three-excitation sector of the Hilbert space. Our theoretical model considers three qubits or quantum emitters confined to a triangle-shaped multimode optical cavity. 

Here, we emphasize that the study of tripartite entanglement is not merely a trivial extension of the single-photon and two-photon bipartite entanglement problem. However, by studying the tripartite case, we enter the much more complex and challenging issue of multipartite entanglement, where both quantification and application of entanglement as a resource have garnered significant interest in recent years \cite{walter2016multipartite}. In particular, tripartite entanglement offers the potential for richer information processing capabilities and improved security in quantum communication protocols \cite{ge2023tripartite}. Unlike bipartite entanglement, which is relatively well characterized, tripartite entanglement presents unique challenges and opportunities, necessitating novel methods for its creation, manipulation, and measurement \cite{yuan2013testing, ziane2018generation,peng2021one,lee2008scalable,teh2014criteria, armata2017harvesting}.

There have been different approaches to the study of multipartite entanglement, most focusing on hybrid atom-mode correlations in a cavity, or even mode-mode correlations mediated through atoms \cite{sundaresan2015beyond, faghihi2014tripartite, wickenbrock2013collective}, others have focused on a different problem altogether, that of varying coupling strengths and atom/cavity decay rates \cite{nayak2010environment,alidoosty2013simulation}. In the present work, we quantify the tripartite entanglement among three qubits using the tripartite negativity measure \cite{sabin2008classification}. To this end, we not only discuss the closed system dynamics (as mainly focused in Ref.~\cite{ficek2012effect, gulfam2013creation} for bipartite entanglement), but we also include the excitation loss through the mechanisms of spontaneous emission and cavity leakages. Since such effects cannot be ignored under realistic conditions of open quantum systems, our findings are considered more valuable in the context of the impact of decoherence on entanglement.

As some of the key results for the single-mode problem under the strong-coupling regime, we find Rabi-like oscillatory behavior in both populations and entanglement among the three qubits, without the presence of any retardation, irrespective of the qubit locations. In the multimode problem, we observe the appearance of the retardation effect (one depending on the location of qubits and the other on the cavity length) and the entanglement collapse and revival pattern, which can be controlled by the locations at which qubits reside in the multimode cavity. Furthermore, we find that, under the strong coupling regime of cQED, the single-mode case always results in the highest value of tripartite entanglement compared to the multimode problem. We also notice that our generated tripartite qubit state overlaps with a genuine tripartite entangled state (Greenberger-Horne-Zeilinger or GHZ state \cite{greenberger1989going}), and we find wide regions of time and strong coupling regime where this overlap (as quantified by the fidelity) can be considerable. In the past, studies have been conducted on the topic of multimode cQED at the level of single or double excitations \cite{bougouffa2013effect, gulfam2013creation}. The main novelty of the present work is the study of tripartite entanglement and the inclusion of realistic loss channels on the generated entanglement, which, to the best of our knowledge, has not been studied before in the present context. 

The remainder of the paper is organized as follows. In Sec.~\ref{sec:II}, we set the stage by providing the theoretical description of our setup, including the model Hamiltonian, modification of the Hamiltonian due to losses, the three-excitation quantum state, and the tripartite entanglement measure utilized for the quantification of entanglement among three qubits. In Sec.~\ref{sec:III} we present our numerical results for tripartite entanglement for single- and multimode problems under different conditions of atomic locations, loss mechanisms, and overlap with the GHZ states. Finally, in Sec.~\ref{sec:IV}, we summarize the key takeaways from this work.

%%===================================================%%
%%      Sec.II  Theoretical Description              %%
%%===================================================%%
\section{\label{sec:II} Theoretical description}

\begin{figure}
\includegraphics[width=2.75in, height=2.9in]{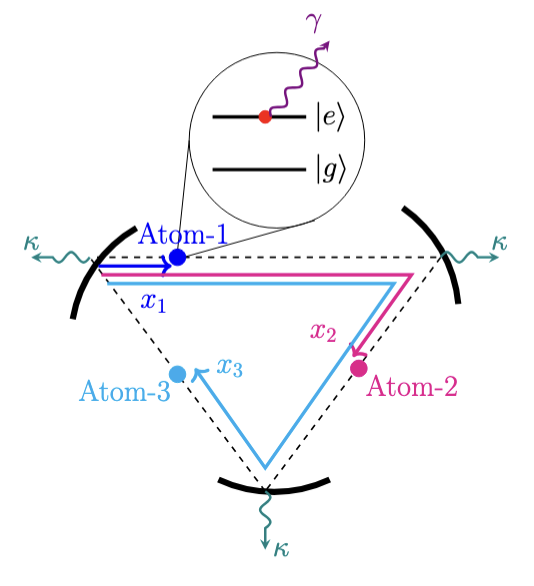} 
\captionsetup{
format=plain,
margin=1em,
justification=raggedright,
singlelinecheck=false
}
\caption{(Color online) Illustration of a multimode triangle cavity with three two-level identical atoms (with ground state $\ket{g}$ and excited state $\ket{e}$) trapped at locations $x_1$, $x_2$, and $x_3$, with a total length of the cavity being $L$. The spontaneous emission and cavity leakage rates are labeled as $\gamma$ and $\kappa$, respectively. For simplicity, we have assumed that the cavity leakage rate for all mirrors is the same. Similarly, the spontaneous emission rate for all atoms has been supposed to be identical. All three atoms are initially excited and unentangled. We are interested in investigating the formation and dynamics of tripartite entanglement among the qubits due to a cavity mode-mediated interaction between them.}\label{Fig1}
\end{figure}
%%%%%%%%%%%%%%%%%%%%%%%%%%%%%%%%%%%%%%%%%%%%%%%%%%%%%%%%%%%%%%%%%%%
\subsection{Model Hamiltonian}
As shown in Fig.~\ref{Fig1}, our model consists of three identical two-level atoms (quantum emitters or qubits) trapped inside a multimode triangular-shaped cavity. The complete Hamiltonian $\hat{\mathscr{H}}$ of the system can be decomposed into three pieces:
\begin{align}\label{TotHam}
\hat{\mathscr{H}} = \hat{\mathscr{H}}_a + \hat{\mathscr{H}}_f + \hat{\mathscr{H}}_{int},
\end{align}
where $\hat{\mathscr{H}}_a$ is the free Hamiltonian for the qubits, $\hat{\mathscr{H}}_f$ is the free Hamiltonian of the quantized multimode electromagnetic field in the cavity, and $\hat{\mathscr{H}}_{int}$ is the Hamiltonian describing the interaction between the qubits and the cavity field. The free Hamiltonian for the qubits can be expressed as
\begin{align}\label{AtomHam}
\hat{\mathscr{H}}_a = \sum_{j=1}^3\hbar\omega_{eg_{j}}\hat{\sigma}^\dagger_{j}\hat{\sigma}_j,
\end{align}
where $\omega_{eg_{j}}$ represents the transition frequency of the $j^{th}$ qubit with $\hat\sigma^\dagger_j:=\ket{e}_j\bra{g}$ and $\hat\sigma_j:=\ket{g}_j\bra{e}$ representing the $j$th qubit raising and lowering operators, respectively. These qubit operators follow the standard Fermionic anti-commutation relation: $\lbrace \hat{\sigma}_j,\hat{\sigma}^\dagger_k\rbrace=\delta_{jk}$, $\forall j=1,2,3$ and $k=1,2,3$.

%The field Hamiltonian
Next, the Hamiltonian of a multimode cavity field can be expressed as
\begin{align}\label{FieldHam}
\mathscr{\hat{H}}_f = \sum_n \hbar\omega_n\hat{a}_n^\dagger \hat{a}_n.
\end{align}
Here, $\hat{a}_n(\hat{a}^\dagger_n)$ represents the photonic annihilation (creation) operator in the $n$th optical mode. The ladder operators for the cavity field follow the standard bosonic commutation relation $\left[\hat{a}_n, \hat{a}^\dagger_{n^{'}}\right] = \delta_{n n^{'}}$. The discretizations of the cavity mode frequency originate from periodic boundary conditions, which give $\omega_n=2\pi n c/L$, where $L$ represents the length of the cavity, $c$ being the speed of light, and $ n=1,2,3,..., N_m$. Therein, $n=1$ can be regarded as the fundamental mode number and $N_m$ represents the highest-order harmonic.  

%The interaction Hamiltonian
For a multi-mode cQED problem, the atom-field interaction Hamiltonian can be written as
\begin{equation}
\hat{\mathscr{H}}_{int}=i\hbar\sum_{j=1}^3\sum_n \bigg(\mathscr{G}_n(\vb{x}_j)\hat{\sigma}_j^\dagger \hat{a}_n - \mathscr{G}^\ast_n(\vb{x}_j)\hat{\sigma}_{j}\hat{a}^\dagger_n\bigg).
\end{equation}
Note that the atom-field interaction parameter $\mathscr{G}_n({\vb{x}_j})$ (which is also sometimes called the position-dependent Rabi frequency \cite{bougouffa2013effect}) here depends both on the mode number $n$ and the location of the $j$th atom $\vb{x}_j$. It is defined through the expression
\begin{align}\label{kvalues}
   \mathscr{G}_n({\vb{x}_j}) = \sqrt{\frac{\omega_n}{2\hbar\epsilon_0 L}} \left(\vb{d}_j\cdot \hat{e}_p\right)e^{i{\vb k}_n\cdot{\vb x}_j},
\end{align}
where $\epsilon_0$ is the permittivity of the free space, ${\vb d}_j$ is the $j$th qubit's dipole transition matrix element, $\hat{e}_p$ is a unit vector representing the polarization of the cavity field in some $p$th direction and ${\vb k}_n$ is the wave number vector. 

%%%%%%%%%%%%%%%%%%%%%%%%%%%%%%%%%%%%%%%%%%%%%%%%%%%%%%%%%%%%%%%%%%%
\subsection{Modification of the Hamiltonian due to losses}
For a realistic treatment of the problem, one needs to incorporate loss mechanisms that can model the interaction between the system and its environment. For our problem, there are mainly two sources of such losses, namely, the spontaneous emission from the qubits (with the rate $\gamma$) and the leakage of the photon from any one of the triangle cavity mirrors (with the rate $\kappa$, which we have assumed to be the same for all mirrors). 

The complete treatment of this problem would involve modeling the environment as a multimode quantum harmonic oscillator and solving the problem under the Born-Markov approximation of open quantum systems \cite{breuer2002theory}. For cavity QED setups, such calculations have already been reported by Shen and Fan (see, for instance, Ref.~\cite{shen2009theory1}) as an example. As the final result, Shen and Fan demonstrate that if the Lamb shift is ignored, the coupling of such quantum optical systems with multimode environments essentially results in a modification of the Hamiltonian. In particular, the bare/free frequencies of the qubit and the cavity modes in the total Hamiltonian $\hat{\mathscr{H}}$ are updated with the decay terms incorporated in the following fashion:
\begin{equation}
    \begin{aligned}
    \hat{\mathscr{H}} &= \sum_{j=1}^3\hbar\tilde\omega_{eg}\hat{\sigma}_{j}^\dagger\hat{\sigma}_j + \sum_{n} \hbar\tilde\omega_{n}\hat{a}_{n}^\dagger \hat{a}_{n}\\ &+i\hbar \sum_{j=1}^3\sum_n \left(\mathscr{G}_{n}({\bf x}_j)\hat{\sigma}_j^\dagger \hat{a}_n - h.c.\right)
    \end{aligned}
\end{equation}
where $h.c.$ stands for hermitian conjugate, $\Tilde{\omega}_{eg} := \omega_{eg} - i\gamma/2$ and $\Tilde{\omega}_n := \omega_n - i\kappa/2$. In the next section, we will use this form of the Hamiltonian whenever we wish to discuss the impact of losses on the dynamics of population and entanglement in the system. 
%%%%%%%%%%%%%%%%%%%%%%%%%%%%%%%%%%%%%%%%%%%%%%%%%%%%%%%%%%%%%%%%%%%
\subsection{Three-photon quantum state}
The quantum state of the system under study, restricted to the three-excitation sector of the Hilbert space, can be expressed as
%\begin{align}
%&\ket{\Psi(t)}= \bigg[\mathcal{A}_{123}%(t)\hat{\sigma}^\dagger_1\hat{\sigma}^\dagger_2\hat{\sigma}^\dagger_3 + \sum\limits_{i, \alpha}\sum\limits_{j>i} \mathcal{A}_{ij\alpha}(t)~\hat{\sigma}^\dagger_i\hat{\sigma}^\dagger_j\hat{a}^\dagger_\alpha + \nonumber\\
%&\sum\limits_{j, %\alpha}\sum\limits_{\beta\geq\alpha}\mathcal{A}_{j\alpha\beta}%(t)\hat{\sigma}^\dagger_j\hat{a}^\dagger_\alpha\hat{a}^\dagger_\beta + \sum\limits_\alpha\sum\limits_{\beta\geq\alpha}\sum\limits_{\gamma\geq\beta}\mathcal{A}_{\alpha\beta\gamma}(t)\hat{a}^\dagger_\alpha\hat{a}^\dagger_\beta\hat{a}^\dagger_\gamma\bigg]\ket{\varnothing},
%\end{align}

\begin{align}
    &\ket{\Psi(t)} = \bigg[\mathcal{A}_{123}(t)\hat\sigma^\dagger_1\sigma^\dagger_2\sigma^\dagger_3+ \sum_{i}\sum_{j>i}\sum_{\alpha} \mathcal{A}_{ij\alpha}(t)\sigma^\dagger_j \sigma^\dagger_i \hat{a}^\dagger_\alpha \nonumber\\ 
    &+\sum_{j}\sum_{\alpha} \frac{\mathcal{A}_{j\alpha\alpha}(t)}{\sqrt{2!}}\hat\sigma_j^\dagger \hat a_\alpha^\dagger \hat a_\alpha^\dagger +\sum_{j}\sum_{\alpha}\sum_{\beta> \alpha} \mathcal{A}_{j\alpha\beta}(t)\hat\sigma_j^\dagger \hat a_\alpha^\dagger \hat a_\beta^\dagger \nonumber\\& +
    \sum_{\alpha}\sum_{\beta> \alpha}\frac{\mathcal{A}_{\alpha\alpha\beta}(t)}{\sqrt{2!}}\hat a_\alpha^\dagger\hat a_\alpha^\dagger\hat a_\beta^\dagger +
    \sum_{\alpha}\sum_{\beta> \alpha}\frac{\mathcal{A}_{\alpha\beta\beta}(t)}{\sqrt{2!}}\hat a_\alpha^\dagger\hat a_\beta^\dagger\hat a_\beta^\dagger + \nonumber\\& 
    \sum_{\alpha}\frac{\mathcal{A}_{\alpha\alpha\alpha}(t)}{\sqrt{3!}}\hat a_\alpha^\dagger\hat a_\alpha^\dagger\hat a_\alpha^\dagger + 
    \sum_{\alpha}\sum_{\beta> \alpha}\sum_{\gamma>\beta} \mathcal{A}_{\alpha\beta\gamma}(t)\hat a_\alpha^\dagger\hat a_\beta^\dagger\hat a_\gamma^\dagger\bigg]\ket{\varnothing},
\end{align}
where $i\in \lbrace 1,2,3 \rbrace$; $j \in \lbrace 1,2,3\rbrace$ are the indices indicating the atom number. Similarly, the $\alpha$, $\beta$, and $\gamma$-indices show the cavity mode number, and $\ket{\varnothing}$ represents the combined ground state of the system in which all qubits are assumed to be in the ground state and no photons are in any mode of the cavity. The probability amplitudes $\mathcal{A}_{123}$, $\mathcal{A}_{i\alpha\alpha}$, $\mathcal{A}_{ij\alpha}$, $\mathcal{A}_{j\alpha\beta}$, $\mathcal{A}_{\alpha\alpha\beta}$, $\mathcal{A}_{\alpha\beta\beta}$, $\mathcal{A}_{\alpha\alpha\alpha}$, and $\mathcal{A}_{\alpha\beta\gamma}$, respectively, represent the cases when all qubits are excited and there are no photons in the cavity field, one atom excited and two photons in the $\alpha$th mode, two atoms excited and one photon in the field, one atom excited and two photons in the field (one in the $\alpha$th mode and other in the $\beta$th mode), and all three atoms in their ground state and all photons in the field - two in the $\alpha$th mode and one in the $\beta$th mode, two in the $\beta$th mode and one in the $\alpha$th mode, all three photons in the same $\alpha$th mode, and all three photons in different modes. The conditions on indices appearing in some summations (for example, $j>i$, $\beta>\alpha$, etc.) are applied to avoid double-counting terms. 

Next, by inserting the total Hamiltonian and the three-excitation state mentioned above into the time-dependent Schrödinger equation $i\hbar\partial_t\Psi(t)=\hat{\mathscr{H}}\ket{\Psi(t)}$, we obtain the equations of motion obeyed by the probability amplitudes. We direct the interested reader to Appendix A and Sec.~\ref{sec:IIIA}, where we report these differential equations for general and single-mode problems, respectively. 

%%%%%%%%%%%%%%%%%%%%%%%%%%%%%%%%%%%%%%%%%%%%%%%%%%%%%%%%%%%%
\subsection{Tripartite entanglement measure}
The tripartite entanglement, which is the entanglement shared among three quantum systems, offers intriguing possibilities in quantum computation and quantum information processing \cite{m2019tripartite, zhang2024entanglement}. One of the key challenges in studying open multipartite quantum systems is to find a robust measure of genuine quantum entanglement \cite{ma2024multipartite, guo2022genuine}. For example, Wotter's concurrence measure \cite{wootters1998entanglement} applicable for the bipartite case cannot be straightforwardly extended to multipartite quantum systems. For this work, we have selected tripartite negativity \cite{kumar2022tripartite, sabin2008classification} as our entanglement measure to quantify genuine entanglement between our three qubits trapped inside the triangle optical cavity. 

To calculate tripartite negativity, we first constructed the full density matrix for our qubit-cavity system $\hat{\rho}(t) = \ket{\Psi(t)}\bra{\Psi(t)}$. Next, since we are interested in quantifying the entanglement between qubits, we took a partial trace with respect to the cavity modes' degrees of freedom, thus finding the qubit or atom density matrix i.e. $\hat{\rho}_a(t) = {\rm tr}_{modes}\lbrace\hat{\rho}(t)\rbrace$. We then introduce the tripartite negativity between the three qubits as
\begin{equation}
    \mathcal{N}(\hat{\rho}_a) = \Big[\mathcal{N}(\hat{\rho}_a^{TA})\cdot \mathcal{N}(\hat{\rho}_a^{TB})\cdot \mathcal{N}(\hat{\rho}_a^{TC})\Big]^{1/3},
\end{equation}
where $\hat{\rho}_a^{T\alpha}$ represents the partial transpose of $\hat{\rho}_a$ with respect to the subsystem $\alpha$ and $\mathcal{N}(\hat{\rho}_a^{T\alpha})$ is the corresponding negativity \cite{sabin2008classification}. The matrix elements of $\hat{\rho}_r^{T\alpha}$ are given by: $\bra{m_\alpha,n_\beta}\hat{\rho}_a^{T\alpha}\ket{p_\alpha,q_\beta} = \bra{p_\alpha,n_\beta}\hat{\rho}_a\ket{m_\alpha,q_\beta}$ where $\alpha = A,B,C$ and $\beta = BC, AC, AB$ with the negativity of a subsystem $\alpha$ being defined as
\begin{equation}
    \mathcal{N}(\hat{\rho}_a^{T\alpha}) = \sum_i \left({|\lambda_i|-\lambda_i}\right).
\end{equation}
Here, $\lambda_i$ are the eigenvalues of $\hat{\rho}_a^{T\alpha}$. This essentially
boils down to twice the sum of negative eigenvalues of the partial transposed density matrix. We bring this to the reader's attention that this differs from the original definition of negativity, as reported in Ref.~\cite{vidal2002computable}, by a factor of 2. The last two equations are the form of the negativity that we will use in our numerical results presented in the next section.

%%===================================================%%
%%                  Sec. III Results                 %%
%%===================================================%%
\section{\label{sec:III} Results}
\subsection{\label{sec:IIIA} Single-mode problem}
\begin{figure*}
\centering
\begin{tabular}{cccc}
\includegraphics[width=2.35in, height=1.72in]{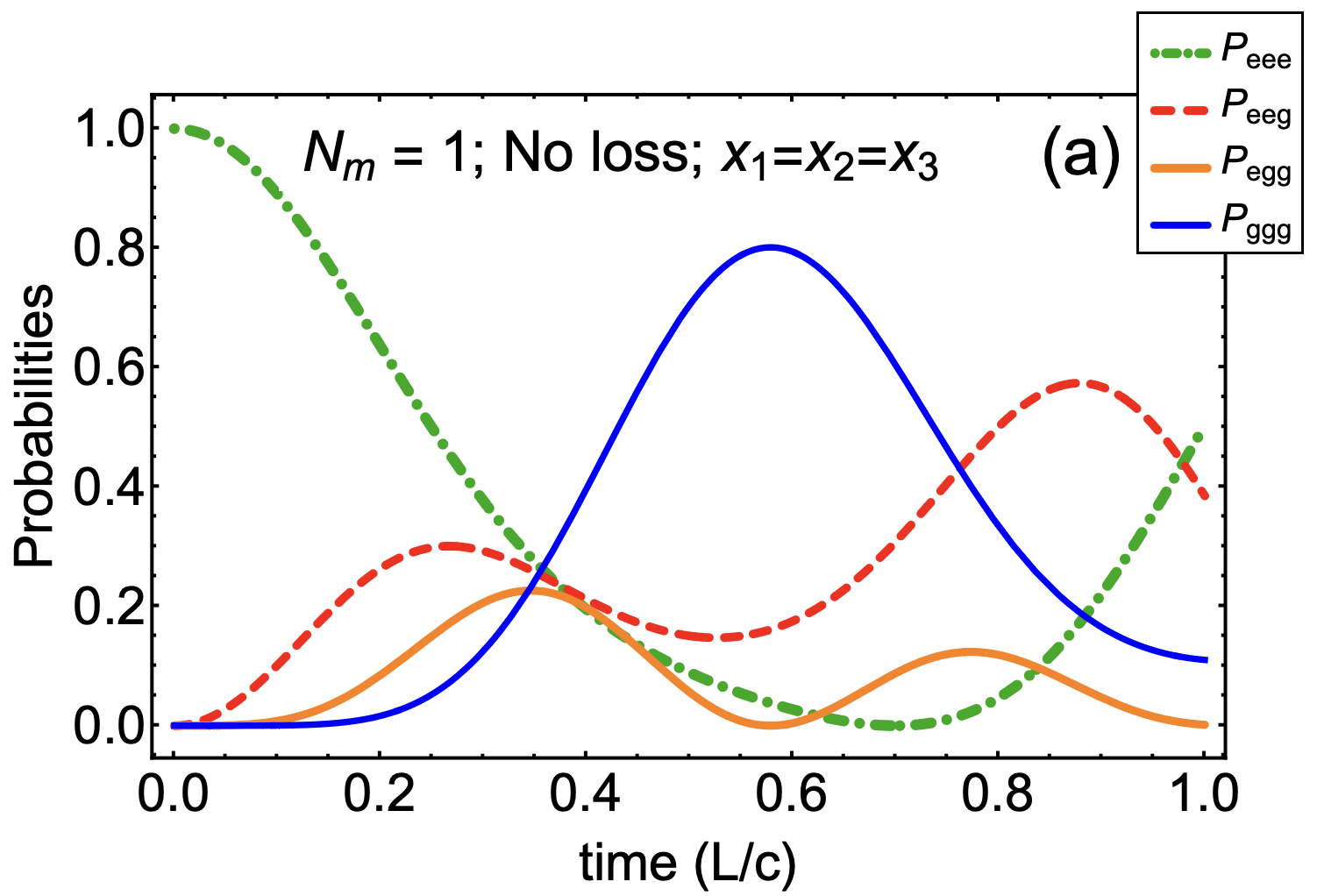} &
\hspace{-1mm}\includegraphics[width=2.25in, height=1.6in]{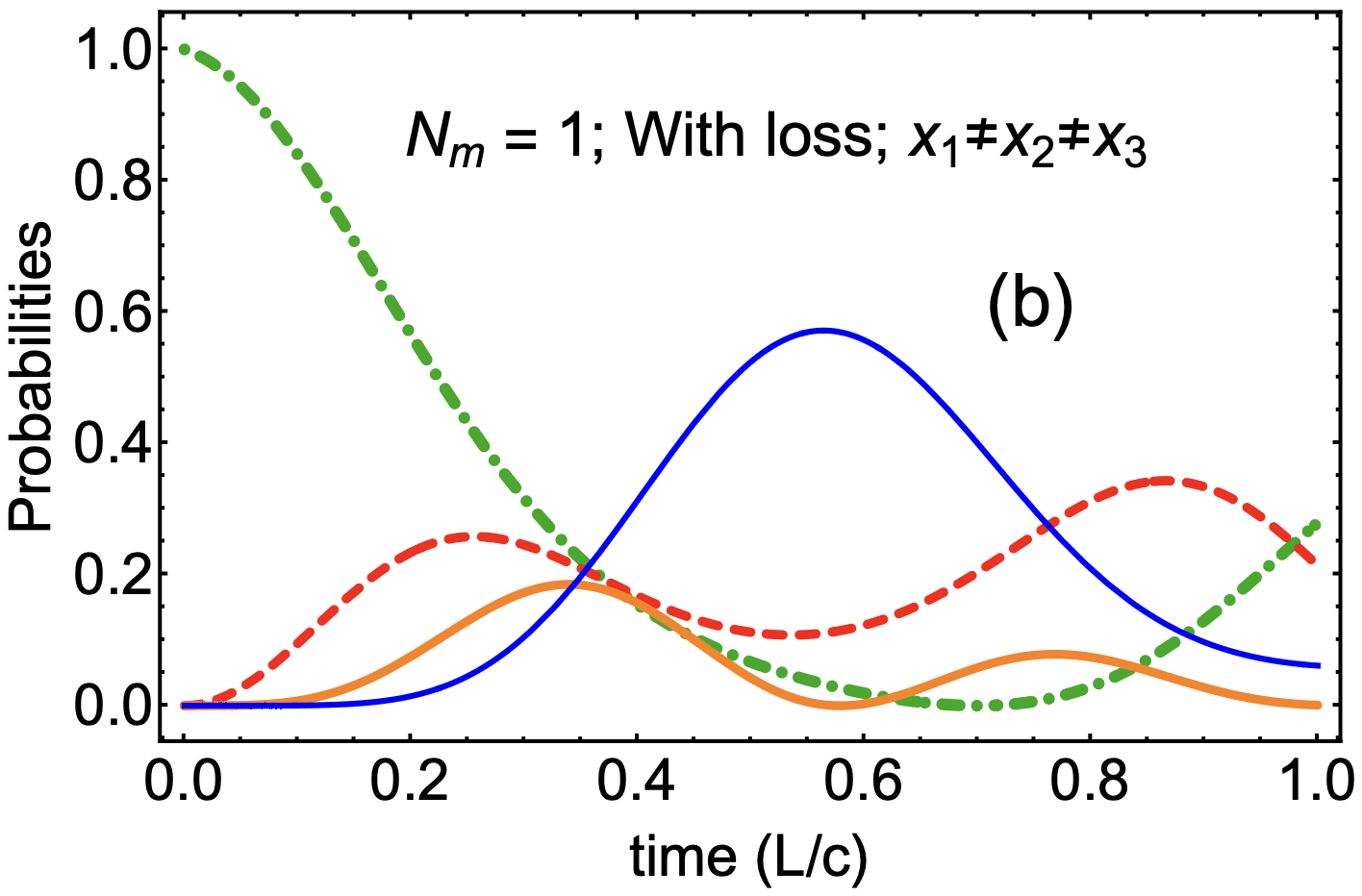} &
\includegraphics[width=2.28in, height=1.64in]{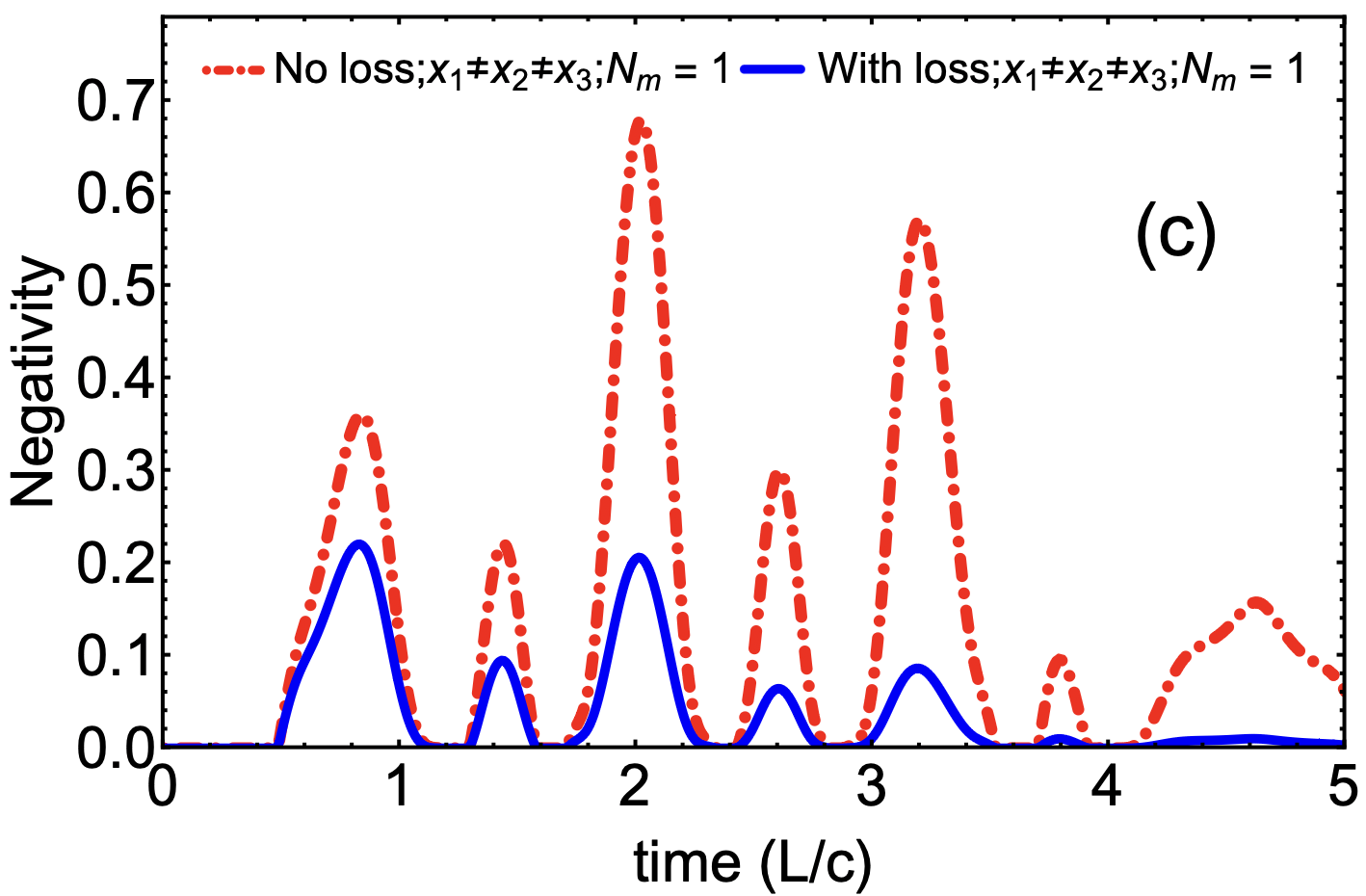} 
\end{tabular}
\begin{tabular}{cccc}
\includegraphics[width=2.25in, height=1.6in]{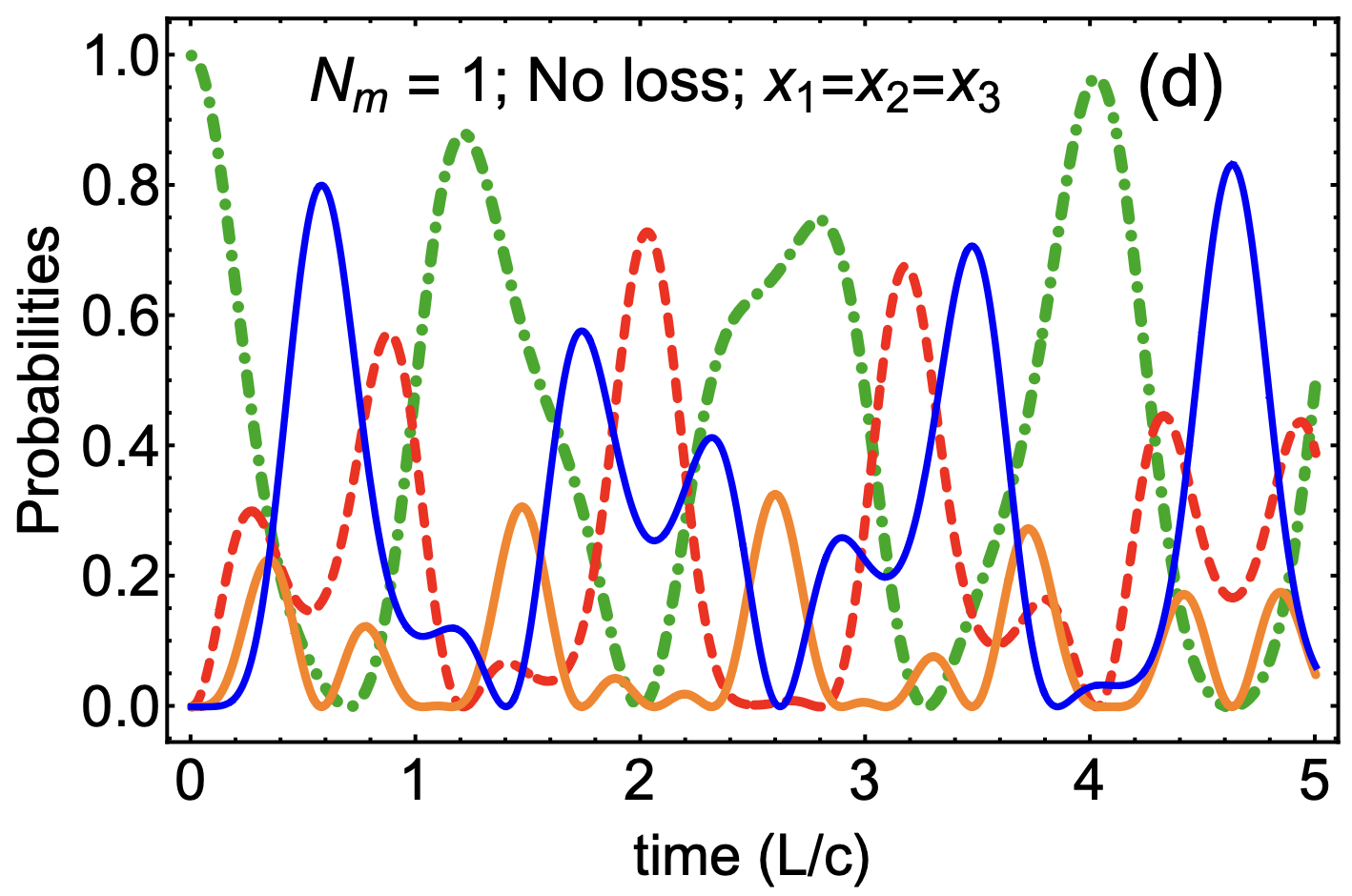} &
\includegraphics[width=2.25in, height=1.6in]{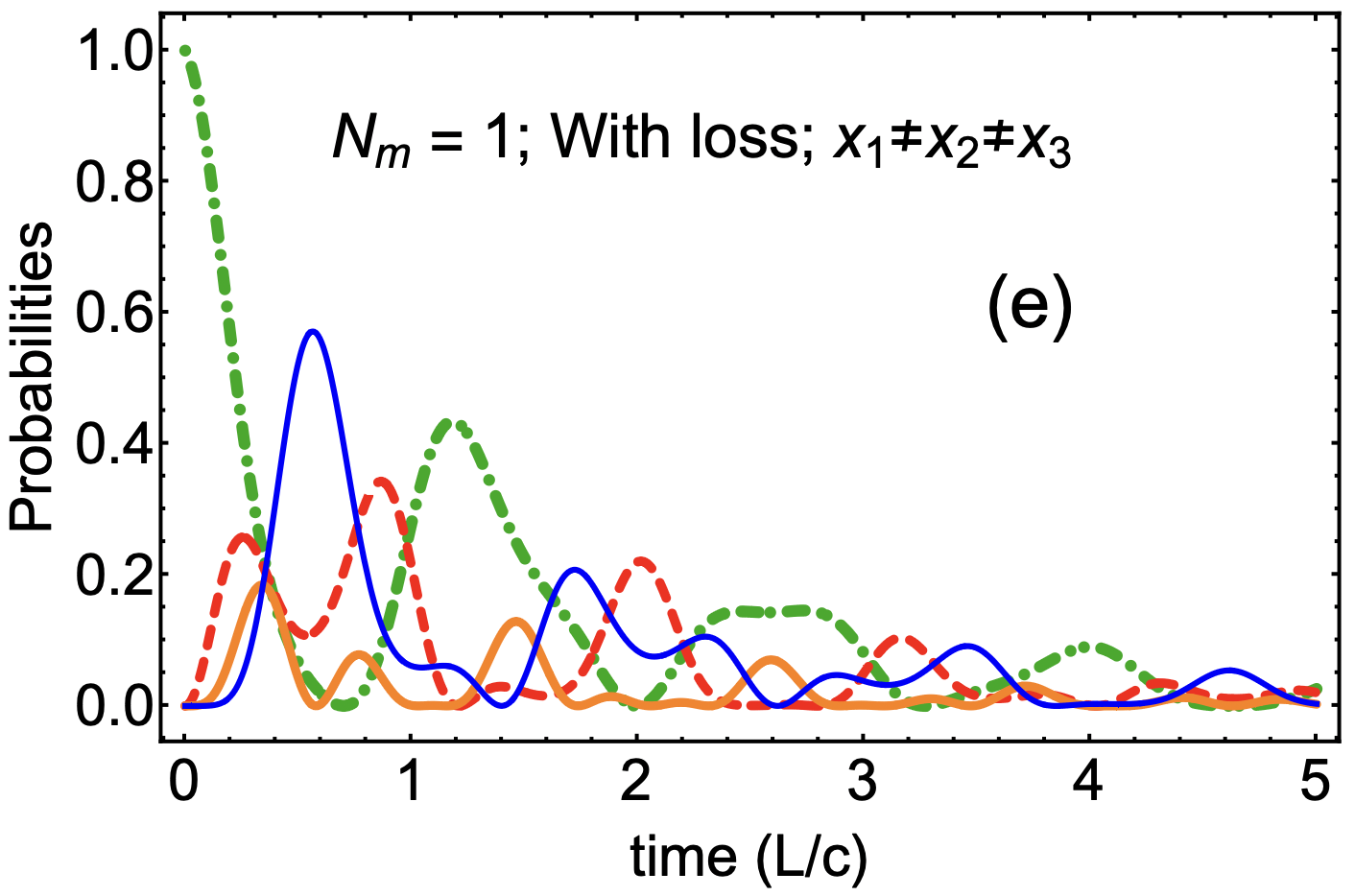}  
\end{tabular}
\captionsetup{
format=plain,
margin=1em,
justification=raggedright,
singlelinecheck=false
}
\caption{(Color online) Plots describing the temporal evolution of probabilities and tripartite negativity for the single-mode problem. (a) No loss case with all qubits at the same location. (d) Same scenario as (a) but with the extended time scale up to $t=5L/c$, where $c$ is the speed of light. (b) With loss case with qubits at different locations. (e) Same situation as (b) but with the extended time scale. (c) Tripartite negativity among three qubits residing at different locations in our triangle cavity. We have compared no-loss and loss cases in this plot. In all plots, we have used the magnitude of qubit-cavity coupling rate $|\mathscr{G}|/(2\pi)= 0.314\Omega_R$ (assumed to be identical for all three qubits with $\Omega_R$ being the vacuum Rabi frequency). For the loss case we have set $\kappa=0.1|\mathscr{G}|$ and $\gamma=0.1|\mathscr{G}|$. When qubits are separated, we have set the location of the first, second, and third qubits to be $x_1=0$, $x_2=L/3$, and $x_3=2L/3$, respectively. $L$ here indicated the total path length of our triangle cavity which we have set $L=994.28\lambda_{eg}$ with $\lambda_{eg}$ (being the length scale of the problem) is the wavelength corresponding to qubit transition frequency $\omega_{eg}$ (again all qubits are assumed to be identical).   }
\label{Fig2}
\end{figure*}
To set the stage, we begin by discussing the simplest case in our model, where a single mode mediates the interaction between the three qubits (we refer to that mode as the $\alpha$th mode). For this problem, according to the time-dependent Schrödinger equation, the evolution of the probability amplitudes is governed by the following set of differential equations:
\allowdisplaybreaks
\begin{subequations}
    \begin{align}
        &\dot{\mathcal{A}}_{123}(t) = -(3\gamma/2) \mathcal{A}_{123}(t)+\sum_i\sum_{j>i} g_{\alpha i}\mathcal{A}_{ij\alpha}(t),\\
        &\dot{\mathcal{A}}_{ij\alpha}(t) = -(\gamma+i\tilde\Delta_{\alpha})\mathcal{A}_{ij\alpha}(t) + \sqrt{2}g_{\alpha i}\mathcal{A}_{j\alpha\alpha} \nonumber\\ 
        &~~~~~~~~~~~~~ + \sqrt{2}g_{\alpha j}\mathcal{A}_{i\alpha\alpha} - g^*_{\alpha {k\neq i,j}} \mathcal{A}_{123}(t),\\
        &\dot{\mathcal{A}}_{i\alpha\alpha}(t) = -(\gamma/2+2i\tilde\Delta_{\alpha})\mathcal{A}_{i\alpha\alpha}(t)-2i\tilde\Delta_\alpha \mathcal{A}_{i\alpha\alpha}(t) + \nonumber\\ 
        &\sqrt{3}g_{\alpha 1}\mathcal{A}_{\alpha\alpha\alpha}(t)-\sqrt{2}\bigg(\sum_{j> i}g_{\alpha j} \mathcal{A}_{ij\alpha}(t)+\sum_{j< i}g_{\alpha j} \mathcal{A}_{ji\alpha}(t)\bigg),\\
        & \dot{\mathcal{A}}_{\alpha\alpha\alpha}(t) = -3i\tilde\Delta_\alpha \mathcal{A}_{\alpha\alpha\alpha}(t)-\sqrt{3}\sum_{i=1}^3 g^*_{\alpha i} \mathcal{A}_{i\alpha\alpha(t)}.
    \end{align}
\end{subequations}
Here $\tilde\Delta_\alpha:=\omega_\alpha-\omega_{eg}-i\kappa/2$ and as before $\tilde\omega_{eg}:=\omega_{eg}-i\gamma/2$ and $\tilde\omega_\alpha:=\omega_\alpha-i\kappa/2$. It is worthwhile to mention that for the single-mode case with two atoms and two excitations in the cavity, four amplitudes are needed to describe all possible ways to distribute these two excitations in the system. This problem was studied in detail by Gulfam et al. \cite{gulfam2013creation}. In the present case, with three qubits and a single mode in the three-excitation regime, the solution requires solving eight coupled differential equations, which is an analytically challenging task. Therefore, we proceed with the numerical analysis of the problem. 

In Fig.~\ref{Fig2} we plot the temporal dynamics of populations and qubit-qubit entanglement (as quantified through negativity) in units of $L/c$  for the single-mode problem. In Fig.~\ref{Fig2}(a), we plot the populations when there are no losses included and all qubits are at the same location (which, for practical purposes, means that the inter-qubit separation is extremely small compared to the wavelength of the resonant field). In this part of the figure we have plotted four curves, namely the probability when all qubits are excited $P_{eee}$ (green dotted-dashed curve), two qubits excited $P_{eeg}$ (red dashed curve), one atom is excited $P_{egg}$ (solid orange line), and all qubits in their ground state $P_{ggg}$ (solid blue curve). Fig.~\ref{Fig2}(d) describes the same plot but for an extended time scale (up to $t=5L/c$). Next, in Figs.~\ref{Fig2}(b) and (e), we plot these populations for the case when there is finite loss in the system and qubits are at different locations (see Fig.~\ref{Fig2} caption for the parameters used). Finally, Fig.~\ref{Fig2}(c) is the plot in which we report the tripartite negativity dynamics for the single-mode problem with qubits at different locations and compare the behavior for the no-loss case (red dotted-dashed curve) with the loss case (solid blue curve).

As some of the key findings, we find that in Fig.~\ref{Fig2}(a), we start with our initial condition, that is, all qubits are excited or $P_{eee}=1$. As soon as one or more qubits relax to their ground state, the probabilities $P_{eeg}, P_{egg}, P_{ggg}$ begin to grow, with $P_{egg}$ growing first, and $P_{ggg}$ growing last. We also observe oscillations in these probabilities, which are hallmarks of Rabi oscillations observed in the strong coupling regime of cQED \cite{mercurio2022regimes}. Here, interestingly, we observe the unique possibility of generating superposition states. For example, at $t=0.875L/c$ we find an equal probability of $\sim 20\%$ of finding qubits in the excited state $\ket{eee}$ and in the ground state $\ket{ggg}$ at the same time. From Fig.~\ref{Fig2}(d), we were able to confirm that these oscillatory behaviors extend to longer times with additional later times when different kinds of superposition states could be formed. 

Next, in Figs.~\ref{Fig2}(b) and (e), we plot the population dynamics when the qubits are at different locations ($x_1 = 0, x_2=L/3, x_3=2L/3$) and there are losses included in the model with cooperativity $\mathcal{C}:=g^2/(\kappa\gamma)=100$. Since $\mathcal{C}>1$, we are working in the strong coupling regime here \cite{kroeze2023high}. We notice that the oscillations of the populations resemble those in the no-loss case; however, the maximum values of the probabilities are considerably reduced. This characteristic becomes clear in the long term, when, around $t=5L/c$, all populations reach a value of almost 0.05, while in the no-loss case, some of these probabilities were around 0.8. 

Finally, in Fig.~\ref{Fig2}(c), we report the genuine entanglement (as quantified through tripartite negativity) among three qubits as a function of time. We have plotted the loss and no-loss cases in the same figure for better comparison. We note that the system of three qubits starts with all qubits being excited, which is in an unentangled or separable state. Therefore, negativity initially takes a null value. The same situation lasts until $t\sim 0.5L/c$ (the time taken by the system to generate the entanglement), and after that we observe that the entanglement begins to rise. Given that we are working in the strong coupling regime of cQED, we observe Rabi-like oscillatory behavior extended to entanglement as well, with a clear pattern of collapse and revival \cite{ficek2006dark, xu2010experimental} for tripartite entanglement. The period of these collapses and revivals (as quantified through the separation between two consecutive maxima in negativity) turns out to have a value of almost $2|\mathscr{G}|$. Compared with the loss and non-loss cases, we find that (up to $ t=5L/c$) the maximum value of entanglement achieved for the no-loss case was approximately $68\%$. However, the inclusion of a finite yet small spontaneous emission ($\gamma=0.1|\mathscr{G}|$) and cavity leakage loss ($\kappa=0.1|\mathscr{G}|$) drastically reduces the maximum value of entanglement achieved. In particular, for this case, we can only attain the entanglement $\sim 20\%$ among three qubits. As expected, the no-loss case always poses an upper bound for the entanglement value achieved in the corresponding loss case, where the entanglement eventually dies out in the loss case around $t=5L/c$.

It should be noted that in all plots of Fig.~\ref {Fig2}, we do not observe any retardation effect in the single-mode problem, regardless of the spatial distribution of the qubits. This is due to the absence of qubit-mediated mode-mode interferences in the single-mode problem, a feature also reported for uniexcitation and biexcitation cases in Ref.~\cite{gulfam2013creation}. In the following subsection, we will discuss the multimode problem, revisiting the retardation effects.

%%%%%%%%%%%%%%%%%%%%%%%%%%%%%%%%%%%%%%%%%%%%%%%%%%%%%%%%%%%%%%%
\subsection{\label{sec:IIIB} Multiple-mode problem}
\subsubsection{\it {\bf Scaling of number of amplitudes with number of modes}}
We now turn our attention to the multimode problem. The inclusion of additional modes that can mediate the interaction between the three qubits makes this problem more complex. To see how the number of amplitudes scales with the mode number while keeping the number of qubits and number of excitations fixed, we derived the following formulae: 
\begin{align}
    \mathcal{S} = \sum_{i=0}^\mathcal{E} \binom{\mathscr{A}}{i}\binom{\mathcal{E}-i+N_m-1}{N_m-1}
\end{align}
Here, $\mathcal{S}$ is the number of amplitudes, $\mathscr{A}$ is the number of atoms, $\mathcal{E}$ is the number of excitations, and $N_m$ is the number of modes. For example, for the single-mode problem (discussed in the last subsection), the above formula confirms the presence of four amplitudes. As examples, for a three- or a seven-mode problem, we find that one would require 38 and 190 amplitudes to form a complete set of bases for the combined atom-field Hilbert space. Moving forward, we decided to work with up to $N_m=31$ as we were able to run the resulting numerical routines conveniently with the computational resources at hand.

\subsubsection{\bf Population dynamics and entanglement evolution plots}
\begin{figure*}
\centering
\begin{tabular}{cccc}
\includegraphics[width=2.35in, height=1.7in]{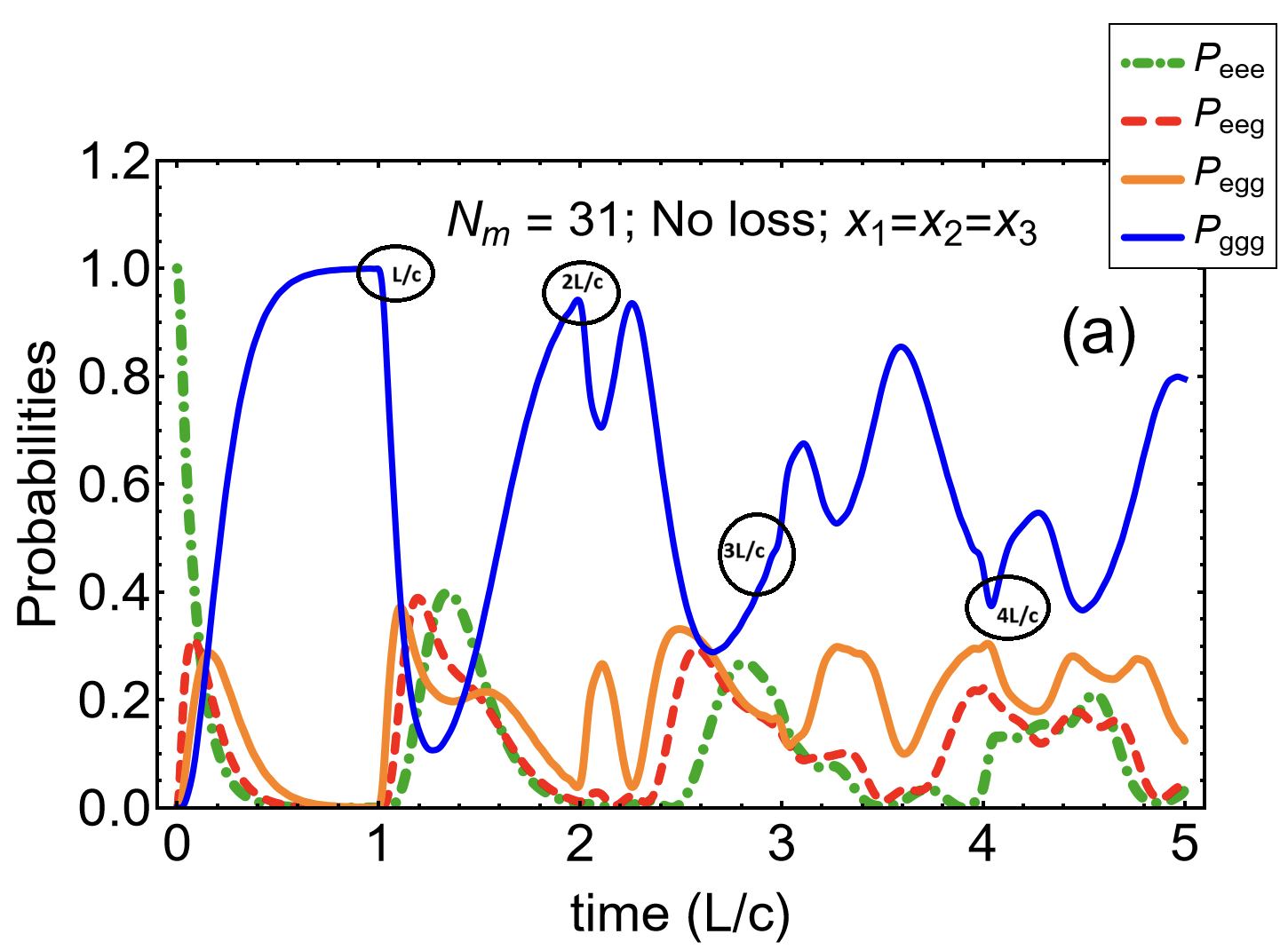} &
\hspace{-1mm}\includegraphics[width=2.25in, height=1.47in]{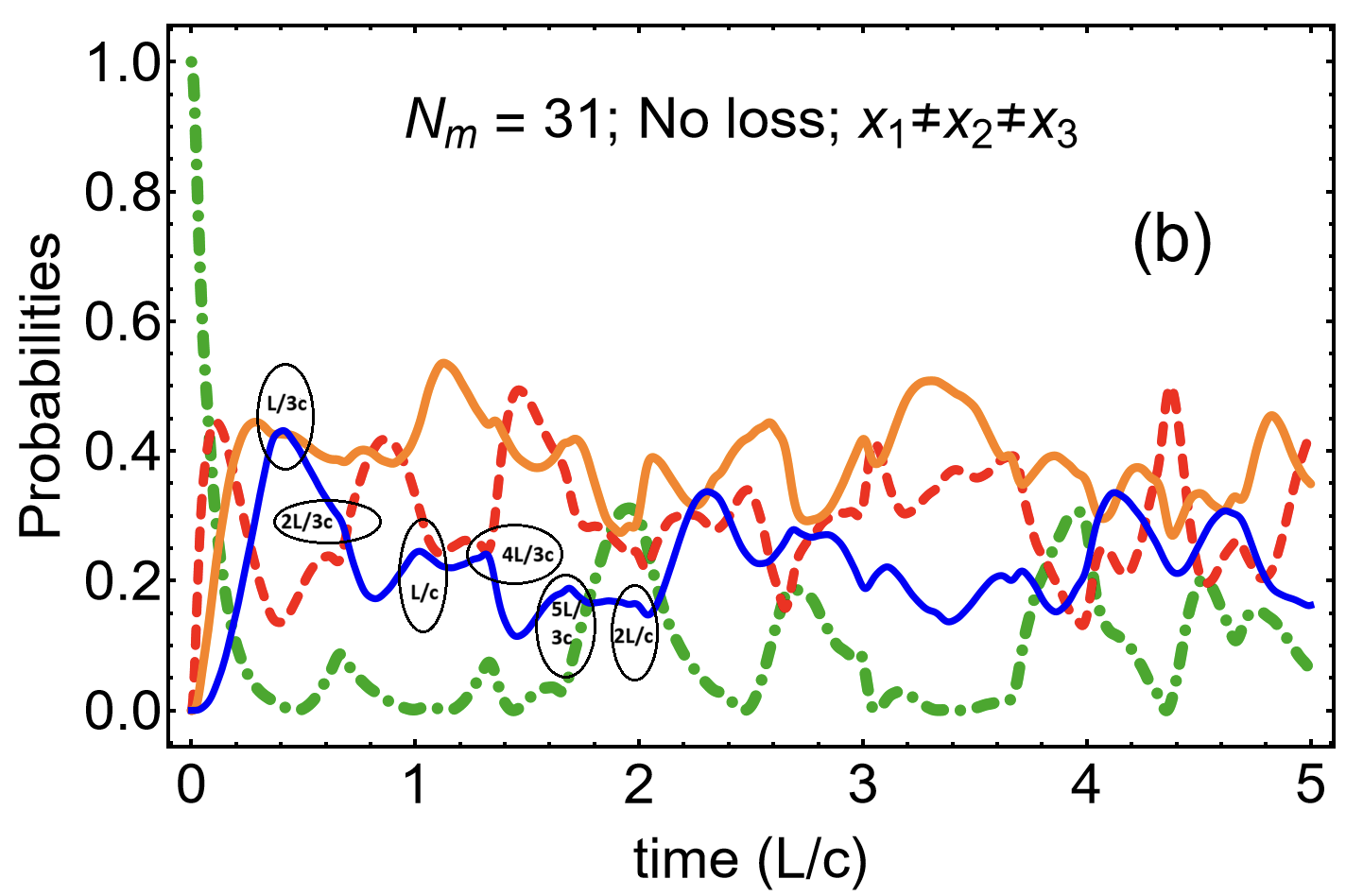} &
\includegraphics[width=2.28in, height=1.5in]{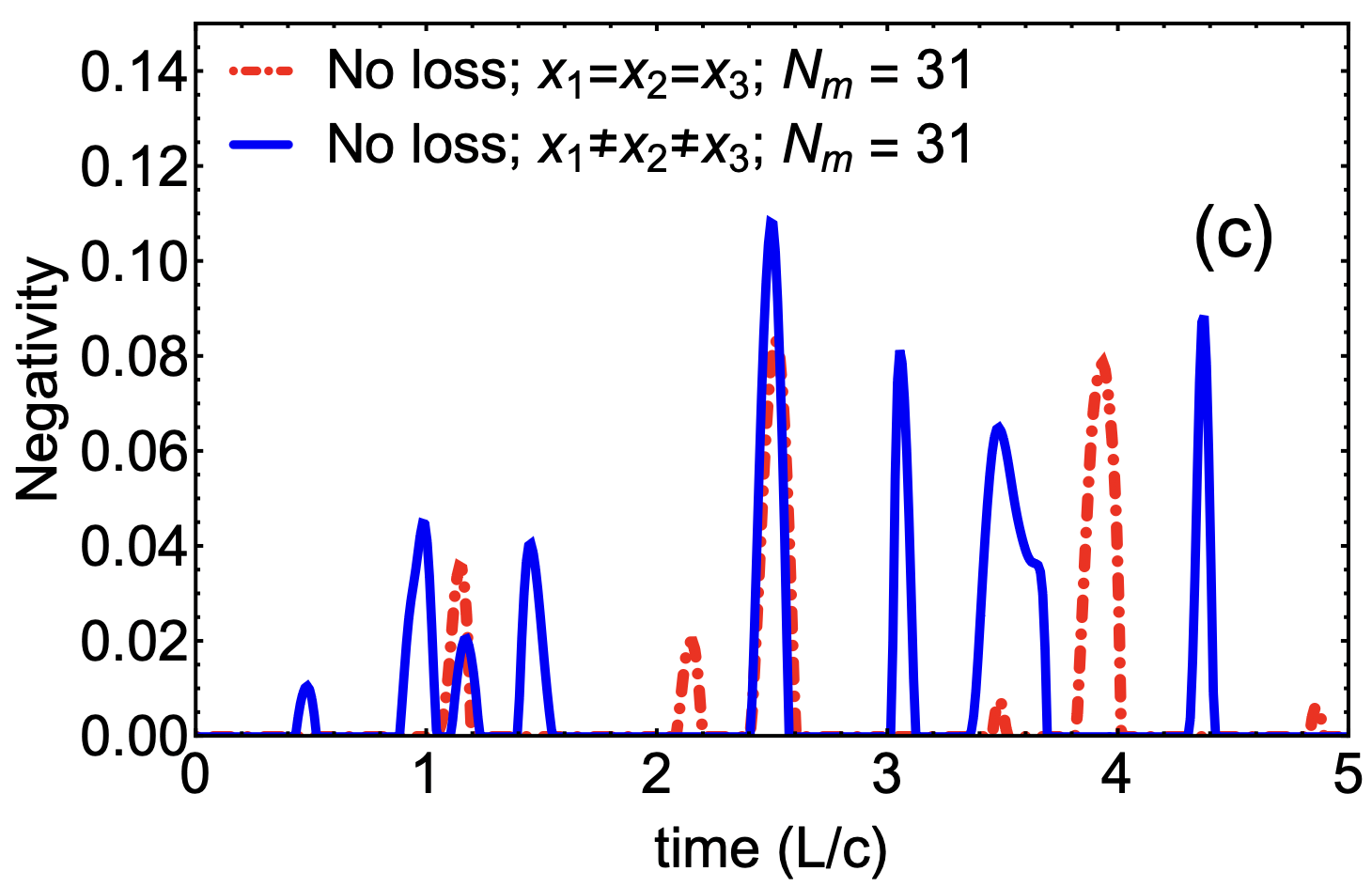} 
\end{tabular}
\begin{tabular}{cccc}
\includegraphics[width=2.25in, height=1.55in]{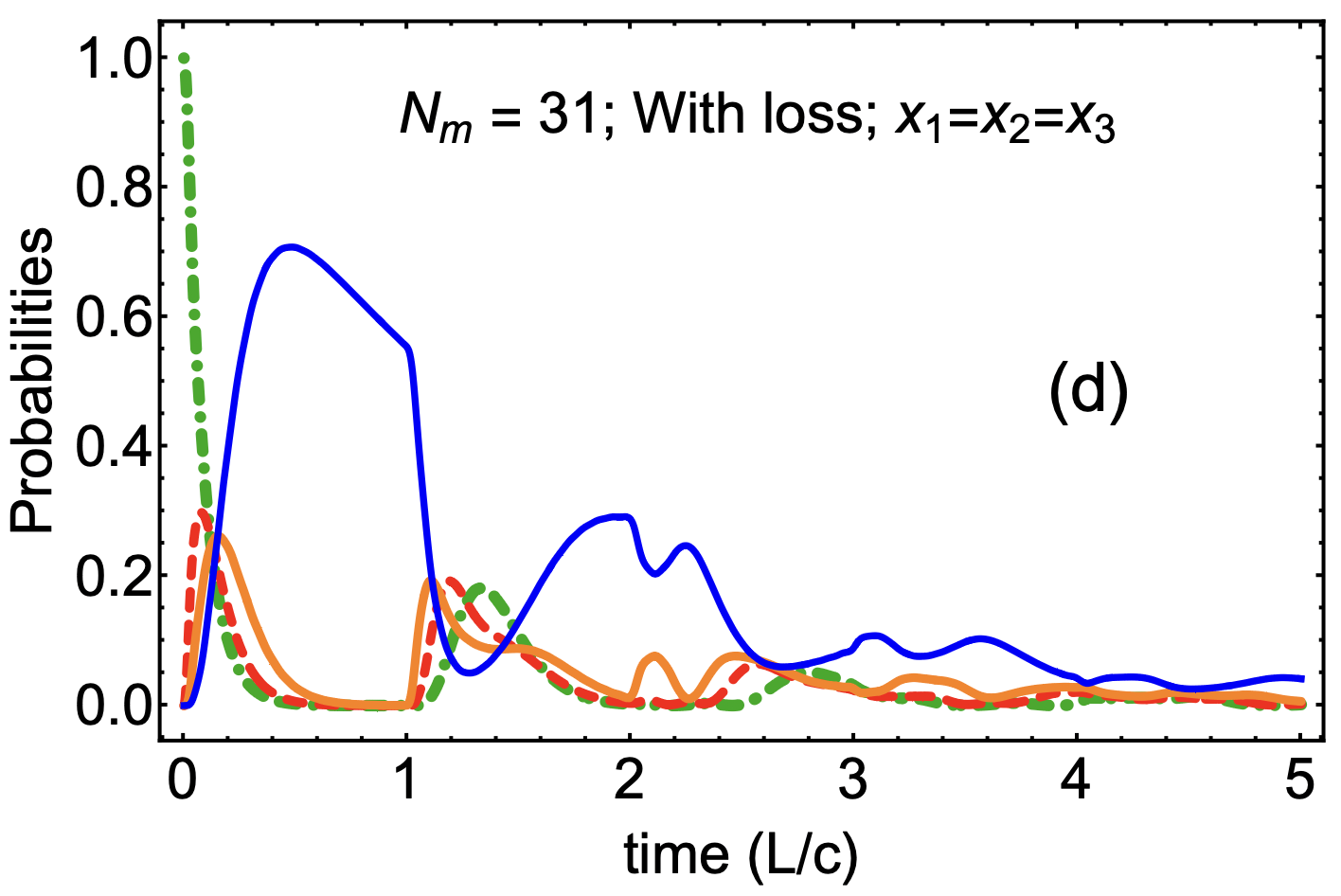} &
\includegraphics[width=2.25in, height=1.55in]{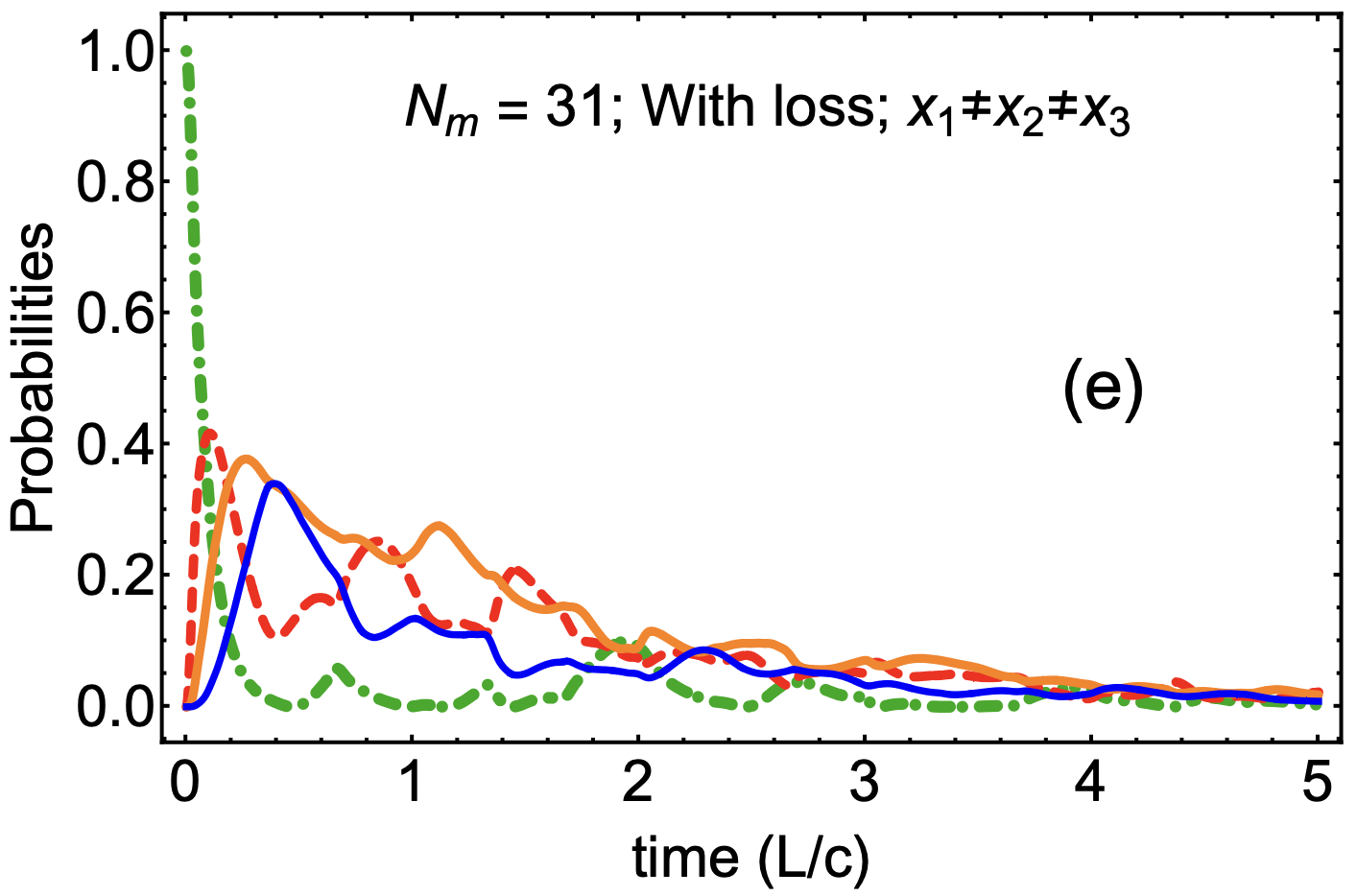}  &
\includegraphics[width=2.25in, height=1.55in]{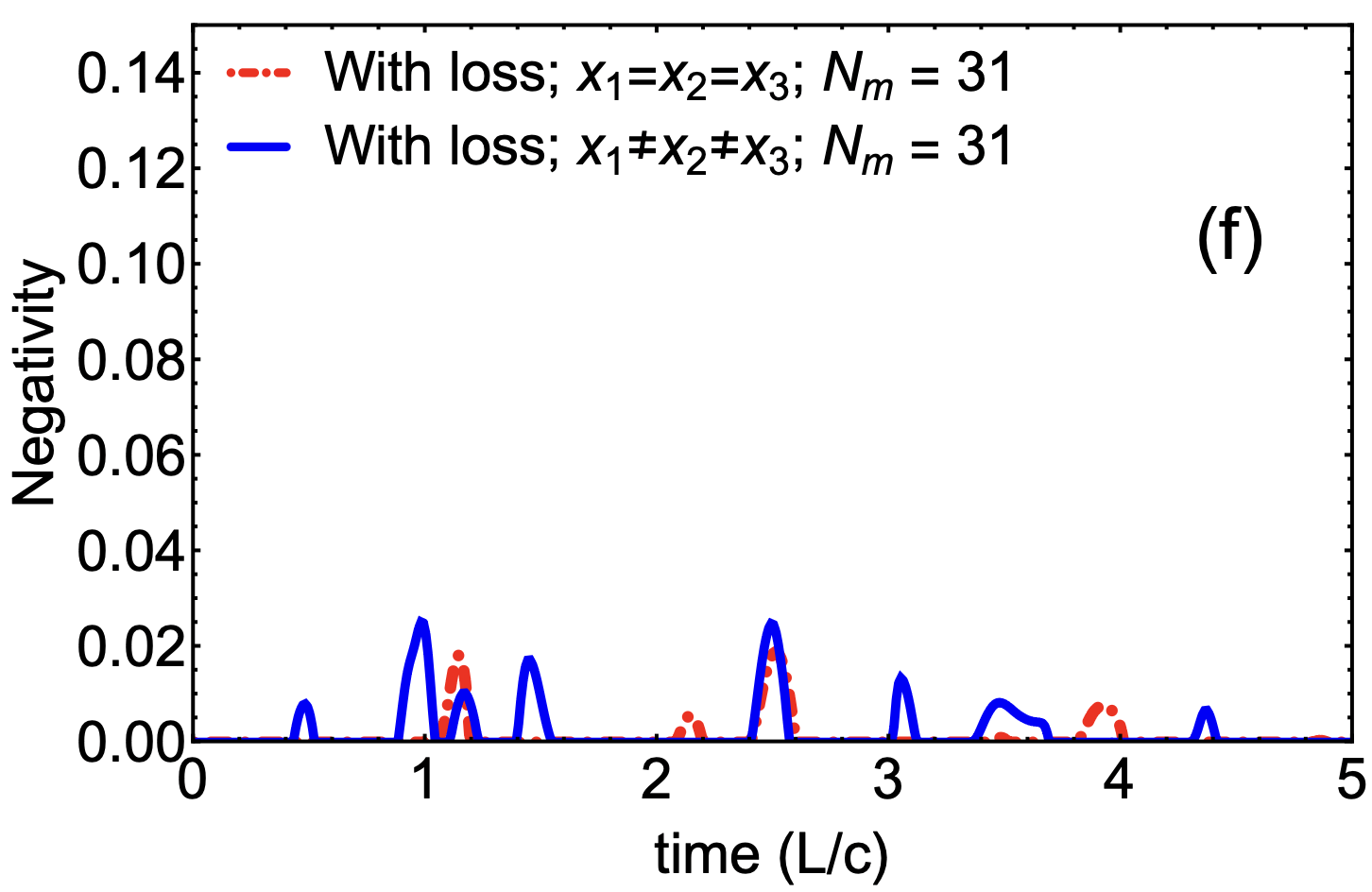} 
\end{tabular}
\captionsetup{
format=plain,
margin=1em,
justification=raggedright,
singlelinecheck=false
}
\caption{(Color Online) Population dynamics and tripartite negativity dynamics plots for the multimode cavity problem. In all the plots we have considered, a triangle cavity with thirty-one independent modes. All modes are assumed to interact with three qubits, with the same coupling rate that we considered for the single-mode problem. In plot (a), we have a no-loss situation with qubits at the same location. Plot (b) is again a no-loss case, but when qubits are at different locations. Plot (c) compares the negativity for these two situations. Finally, plots (d), (e), and (f) are the with-loss versions of plots (a), (b), and (c), respectively. All parameters are the same as used for the single-mode problem of Fig.~\ref{Fig2}. } 
\label{Fig3}
\end{figure*}

We present the first set of plots for the multimode problem in Fig.~\ref{Fig3}. It should be noted that the way we have structured the plots in Fig.~\ref{Fig3} differs slightly from what we did in Fig.~\ref{Fig2}. The reason is that for the multimode case, qubits at the same location and qubits at different locations produced markedly different results due to the interference of multiple modes; however, in the single-mode problem, that was not the case. 

Additionally, for the multimode problem the wavenumbers ${\bf k_n}$ appearing in the qubit-photon coupling rate $\mathscr{G}_n({\bf x}_j)$ (see Eq.~\eqref{kvalues}) and its relationship to $\alpha$th-mode detuning $\Delta_\alpha$ are important to point out here. We notice
\begin{align}
    \Delta_\alpha = \omega_\alpha-\omega_{eg} = 2\pi c\left(\frac{\alpha}{L}-\frac{1}{\lambda_{eg}}\right).
\end{align}
For a linear dispersion case, we know $\Delta_\alpha = c\Delta k_\alpha$. Furthermore, if we use $L=994.28\lambda_{eg}$, we arrive at the following result
\begin{align}
    \Delta k_\alpha=\frac{2\pi}{\lambda_{eg}}\left(\frac{\alpha}{994.28}-1\right).
\end{align}
This yields the separation between any $(\alpha+1)$th and $\alpha$th modes taking the value.
\begin{align}
    \Delta K :=\Delta k_{\alpha+1} - \Delta k_{\alpha} = \frac{2\pi}{\lambda_{eg}(994.28)}.
\end{align}
This is the consecutive mode wavenumber separation which we will use in the numerical results to follow. 

In Fig.~\ref{Fig3}(a), we present our first multimode plot with $N_m=31$ when all three qubits are at the exact location and there are no losses in the system. As discussed in Ref.~\cite{gulfam2013creation} for the single- and double-excitation problems in multimode cQED, in this situation the retardation effects begin to show up because of the time it takes photons to complete one round trip in the optical cavity. We observe the same effect here, where, exactly $t=L/c$, there is a sharp jump in the populations (most visible for the blue solid curve or the $P_{ggg}$ curve). Later, at $t=2L/c, 3L/c,...$, these abrupt jumps tend to become less pronounced due to the presence of three excitations undergoing various processes under the strong coupling regime of cQED. For clarity, we have marked these jump points in Fig.~\ref{Fig3}(a) up to $t=4L/c$ and in Fig.~\ref{Fig3}(b) up to $t=2L/c$ (although these jump points continue at later times as well).   

Next, in Fig.~\ref{Fig3}(b) we report the case same as in Fig.~\ref{Fig3}(a) but now when qubits are at different locations. According to~\cite{gulfam2013creation}, such a situation would include another mechanism of quantum interference leading to retardation effects appearing at places other than $t=2L/c, 3L/c,...$. For the first cycle of photon round trips, these new jumps would be expected to occur at $L\pm x_1 =L$, $L\pm x_2 = 2L/3~\text{or}~4L/3$, $L\pm x_3 = L/3~\text{or}~5L/3$, etc. We observe these retardation effects as new minor jumps/kinks in all curves in Fig.~\ref{Fig3}(b). As expected, when we include losses in Fig.~\ref{Fig3}(d) and Fig.~\ref{Fig3}(e), we find the same curves as found in Fig.~\ref{Fig3}(a) and Fig.~\ref{Fig3}(b), respectively, with the main difference of an overall damping of the oscillations in the populations. 

In the entanglement plot (Fig.~\ref{Fig3}(c)), we make an interesting observation. For the no-loss case, qubits residing at different locations tend to produce on average a higher value of entanglement with the possibility of generating entanglement collapse or revival at times when there was no entanglement present in the corresponding case of when qubits were at the same location (see, for instance, around $t=3L/c$ and $t=9L/(2c)$). Furthermore, we find that by selecting different trapping locations of qubits in this problem, we can manipulate the times at which maximum tripartite entanglement could be revived (plots not shown here). This result can open up new venues for engineering exotic entangled states. Finally, in Fig.~\ref{Fig3}(f) we find that the presence of spontaneous emission and cavity leakage losses drastically reduces the maximum value of negativity achieved by almost a factor of 5 while retaining the pattern of collapse and revival of the entanglement between the qubits. 

%%%%%%%%%%%%%%%%%%%%%%%%%%%%%%%%%%%%%%%%%%%%%%%%%%%%%%%%%%%%%%%%%%%%%%
\subsubsection{\bf Maximum negativity as a function of number of modes}
\begin{figure}
\includegraphics[width=2.9in, height=1.9in]{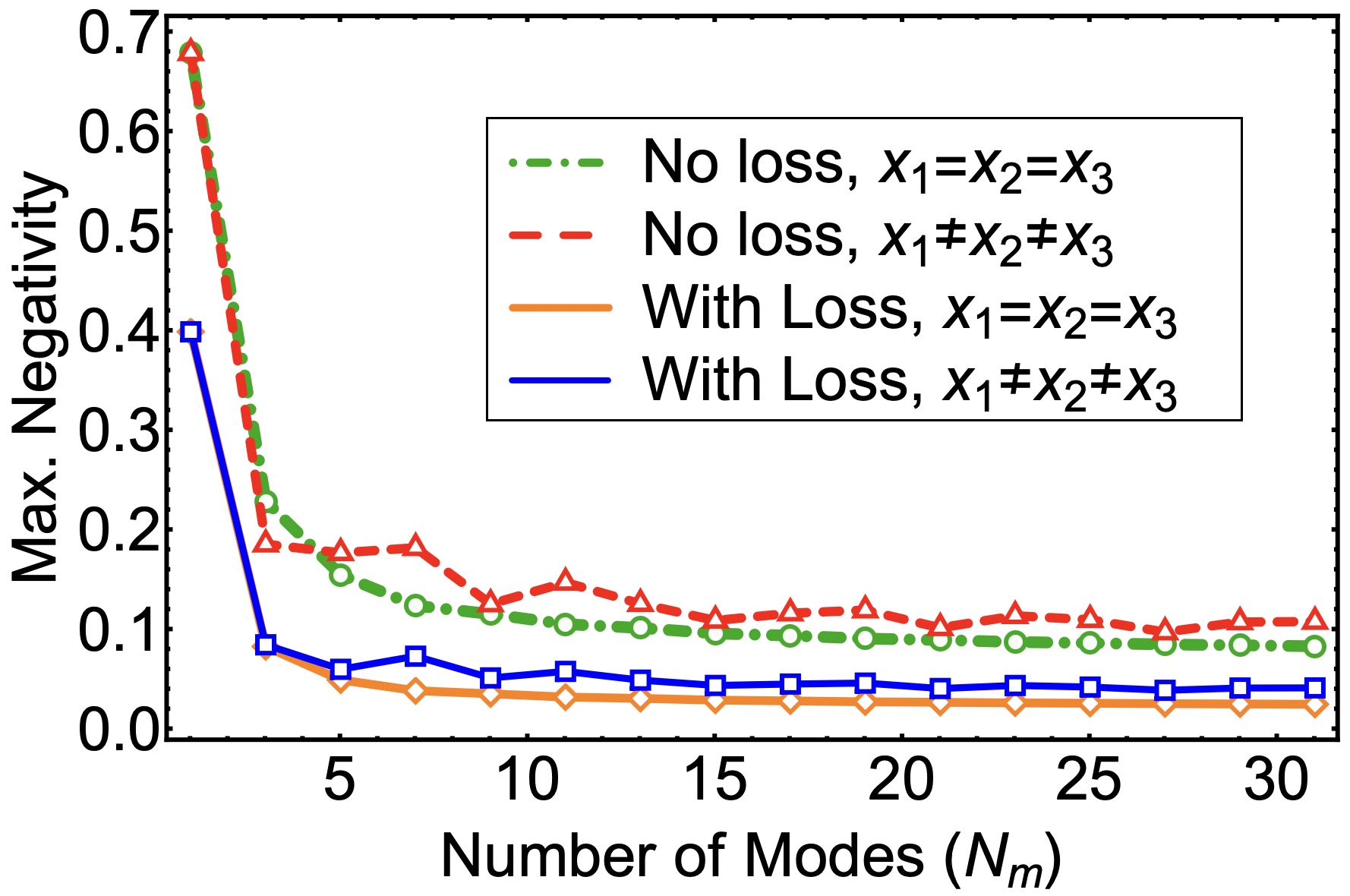} 
\captionsetup{
format=plain,
margin=1em,
justification=raggedright,
singlelinecheck=false
}
\caption{(Color Online) Maximum negativity plot as a function of the number of modes for our three-qubit multimode cQED setup. Four relevant cases of no loss, with loss, qubits at the same location, and qubits at different locations have been presented here. For this plot, we have restricted ourselves to $t=5L/c$ to numerically extract the maximum value of negativity. The remaining parameters are the same as those used in the previous figures.}
\label{Fig4}
\end{figure}

\begin{figure*}
\centering
\begin{tabular}{cccc}
\includegraphics[width=2.35in, height=1.72in]{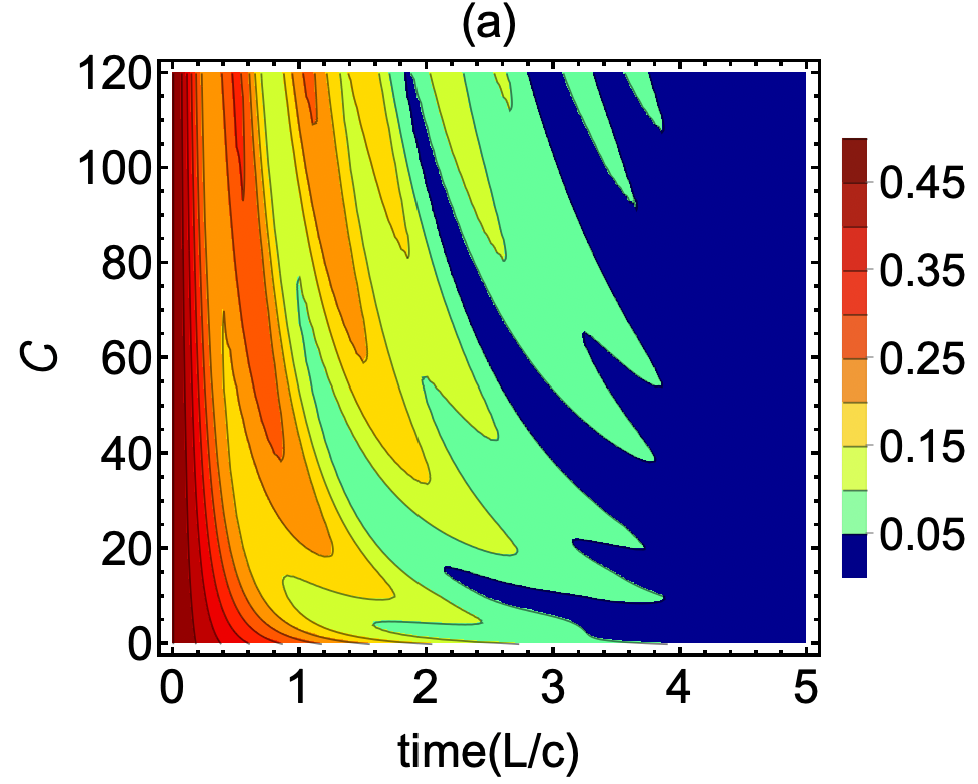} &
\includegraphics[width=2.35in, height=1.72in]{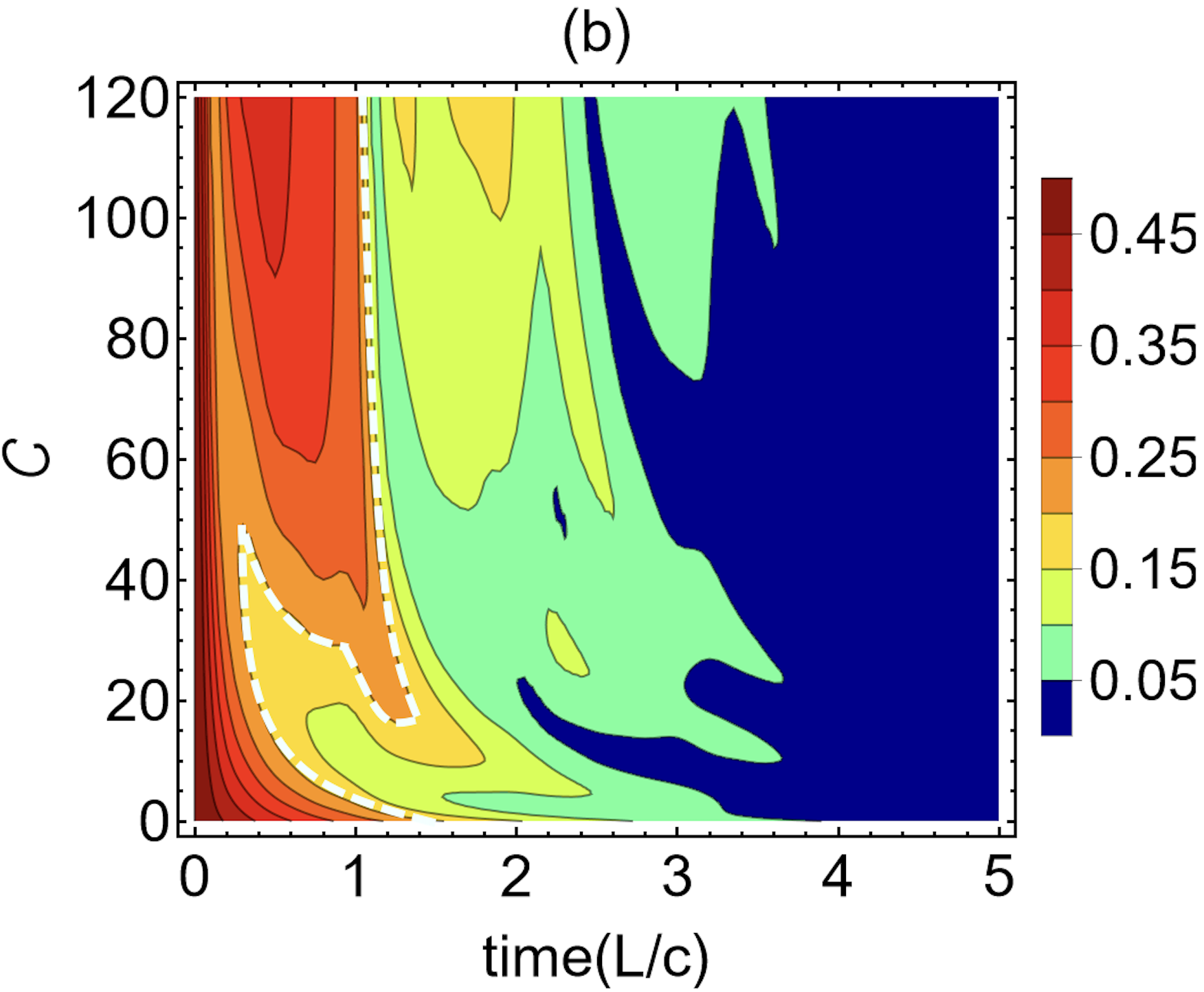} &
\hspace{-5mm}\includegraphics[width=2.35in, height=1.72in]{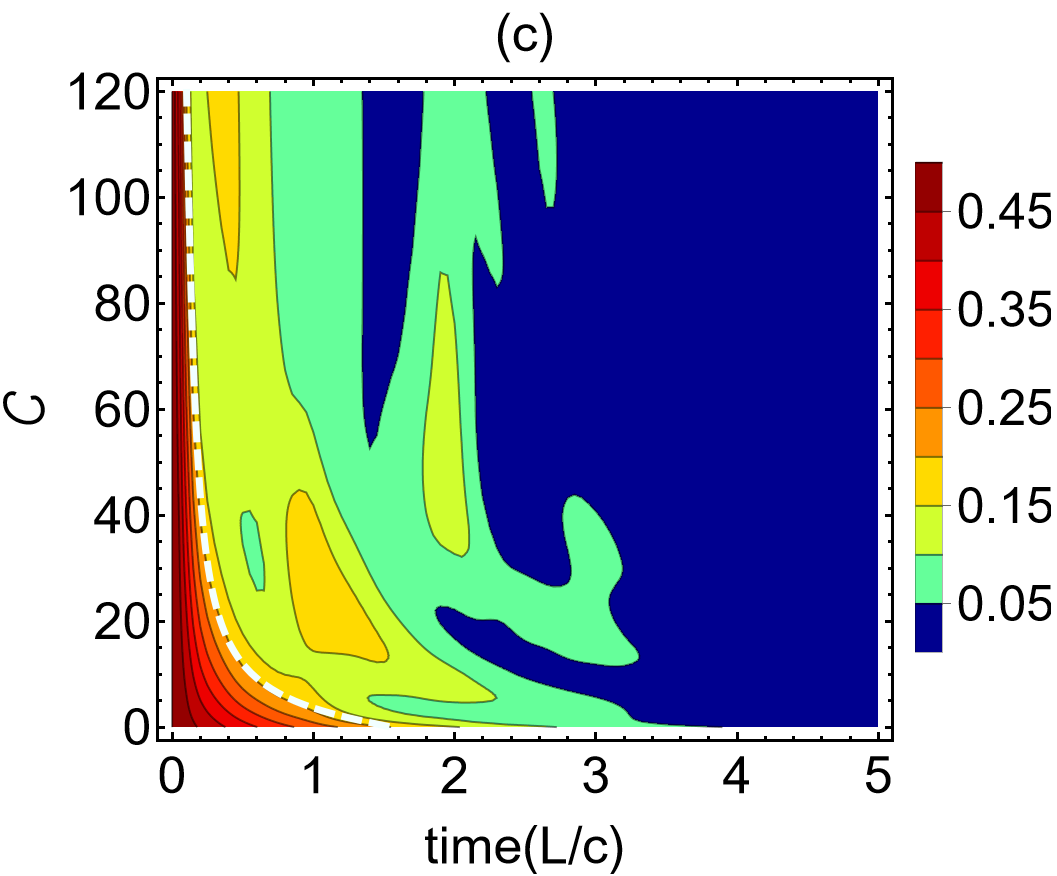} 
\end{tabular}
\captionsetup{
format=plain,
margin=1em,
justification=raggedright,
singlelinecheck=false
}
\caption{(Color Online) Density plots representing the fidelity of our tripartite entangled state benchmarked against the ${\rm GHZ}$ state as a function of time and cooperativity parameter $\mathcal{C}$ of cQED. We have considered three cases of interest here. In plot (a), we consider a single-mode problem, i.e., $N_m=1$, and a case where qubits are located at different positions. The case of seven modes, i.e., $N_m=7$, is plotted in (b), where the qubits remain in the same location. Finally, plot (c) presents the same multimode problem, but with the qubits located at different positions. Again, parameters used are as before.  }
\label{Fig5}
\end{figure*}
For the present problem, a crucial question is how the number of cavity modes affects the generated entanglement. We have discussed this problem from the perspective of the temporal evolution of entanglement in the last figure. Here we present our numerical results concerning the maximum value of entanglement achieved among the qubits as a function of the number of modes in the cavity. To this end, we restricted ourselves to times up to $t=5L/c$ and considered four cases of interest. These cases were selected based on whether there are losses in the system and/or qubits are at the same or different locations, providing a comprehensive view of the conditions that can influence entanglement.

We first notice that, irrespective of the case under discussion in the strong-coupling regime of cQED, the single-mode problem always yields the highest value for maximum entanglement compared to the multi-mode problem. Secondly, as we pass $N_m=10$ for both the no-loss curves (compare red dashed and green dotted dashed curves) and with loss curves (solid blue curve with the orange solid curve), qubits residing at different locations always generated slightly higher maximum entanglement between qubits compared to when qubits were at the same locations.

As expected, the inclusion of losses had a significant impact on the system, reducing the maximum entanglement considerably compared to the case of no loss. This finding underscores the importance of considering losses in our understanding of entanglement, which is one of the novel aspects of our study. Finally, despite being slight oscillations between $5\leq N_m\leq 12$ for the case when qubits were at different locations, we notice that for a large mode number (when $N_m\geq 15$) all maximum negativity curves seem to take an almost constant value, indicating the small-in-magnitude yet stabilized maximum entanglement dependence on the number of optical modes.

%%%%%%%%%%%%%%%%%%%%%%%%%%%%%%%%%%%%%%%%%%%%%%%%%%%%%%%%%%%%%%%%%%%%%%%
\subsubsection{\bf Fidelity with respect to GHZ states}
To gain further insight into the dynamics of tripartite entanglement in the multimode problem, Fig.~\ref{Fig5} reports the fidelity of the tripartite entangled state generated in our setup, using the Greenberger–Horne–Zeilinger state (GHZ state) as our ideal/target state with genuine tripartite entanglement. Furthermore, to study both the dynamics of fidelity and the impact of strong or weak coupling regimes in cQED, we have presented density plots of fidelity as a function of time and the cooperativity parameter $\mathcal{C}$. Note that, so far, in all previous multimode plots, we have used $ N_m = 31$. However, for the sake of simplicity for our numerical calculations in Fig.~\ref{Fig5}, we have restricted ourselves to $N_m=7$ for multi-mode plots. 

In the present case, the three-qubit GHZ state \cite{greenberger1989going} can be expressed as follows
\begin{align}
    \ket{\Psi}_{\rm GHZ} = \frac{1}{\sqrt{2}}\left(\ket{\rm ggg} + \ket{\rm eee}\right).
\end{align}
Since after tracing out the optical modes, our qubit state would be a mixed state (irrespective of losses being included or not), we use the following definition for numerically calculating the fidelity $\mathcal{F}$ of the generated qubit state here with the aforementioned GHZ target state \cite{jozsa1994fidelity}
\begin{align}
    \mathcal{F} = \left({\rm tr}\left\lbrace \sqrt{\sqrt{\hat{\rho}_a}~\hat{\rho}_{\rm GHZ}~\sqrt{\hat{\rho}_a}} \right\rbrace\right)^2, 
\end{align}
where $\hat{\rho}_a$ represents the three-qubit mixed state density operator generated in our setup, while $\hat{\rho}_{\rm GHZ}$ is the density operator constructed from the pure GHZ state mentioned above. 

In Fig.~\ref{Fig5} we plot the numerically calculated fidelity for three cases: (a) single-mode problem with qubits at different locations, (b) multimode problem with qubits at the same position, and (c) multimode problem with qubits at different positions while varying coopertivity $\mathcal{C}$ from 0.005 to 120 with $t$ being varied from $0$ to $5L/c$. As expected in all three plots, we observe that fidelity begins at a value of $50\%$. However, as time passes, for any fixed value of $\mathcal{C}>1$, fidelity begins to show oscillatory behavior in all cases. We attribute these oscillations to Rabi oscillations as a result of the strong coupling regime of cQED. In the single-mode problem, these oscillations tend to survive around $t\sim 4L/c$ for more or less all values of $\mathcal{C}$. In the multimode case, we find that when the qubits were at the same location, a considerably broad region of $\mathcal{F}\sim 0.4$ existed up to $t\sim 1.5L/c$. However, in the corresponding Fig.~\ref{Fig5}(c), such a region shrunk down to a tiny strip for all $\mathcal{C}$ values. To emphasize this point, we have drawn white-dashed contour curves in Fig.~\ref{Fig5}(b) and Fig.~\ref{Fig5}(c) to indicate these regions in the plots. We attribute this trend to the fact that when qubits are far apart and the coupling is strong, the chances of forming an all-qubit excited or all-qubit ground state become challenging as compared to when qubits are at the same location.

%%===================================================%%
%%                 Sec.IV: Conclusions              %%
%%===================================================%%
\section{\label{sec:IV} Summary and Discussion}
In this work, we have studied the problem of tripartite population dynamics and entanglement in single- and multimode cQED setups. Our system consisted of three qubits trapped inside a triangle-shaped cavity with the possibility of losing photons through spontaneous emission and cavity leakage channels. As the main finding of this work, we observed the appearance of retardation effects and the collapse and revival of tripartite negativity for the multimode problem, which is attributed to the quantum interference effect. However, a single-mode problem was capable of producing a higher value of maximum entanglement in the strong-coupling regime. Furthermore, the overlap with the genuine tripartite GHZ state indicates that the multimode problem with qubits at the same location yields vast regions in time and cooperativity, achieving around $40\%$ fidelity. In general, the impact of losses (which was another novel aspect of this work) drastically reduced the generated qubit-qubit entanglement. Overall, despite being a fundamental quantum optical study of qubits coupled to multimode cavities, we find that the location of qubits is a parameter that can be used to engineer retardation in tripartite entanglement, which could have interesting applications in long-distance quantum communication protocols. 

The presence of three excitations, three qubits, and multiple modes offers a rich setup to address additional questions. For example, we explore how a tripartite entangled initial state among qubits (say a GHZ state or a Wolfgang Dür or a W state \cite{dur2000three}) would evolve in time, given the nonlinearities introduced by the strong coupling regime of cQED. Another problem is how to transfer maximally entangled qubit states to the field when there are multiple modes in the cavity. We leave this and similar intriguing problems for future extensions of this work.

%%%%%%%%%%%%%%%%%%%%%%%%%%%%%%%%%%%%%%%%%%%%%%%%%%%%%%%
\acknowledgments
Financial support for this work was provided by the National Science Foundation Grant No. LEAPS-MPS 2212860, and Miami University College of Arts \& Science and Physics Department start-up funding.
%%%%%%%%%%%%%%%%%%%%%%%%%%%%%%%%%%%%%%%%%%%%%%%%%%%%%%%

\vspace{10mm}
%%===================================================%%
%%                Article Appendices                 %%
%%===================================================%%
\setcounter{equation}{0}
\renewcommand\theequation{A.\arabic{equation}}
\section{APPENDIX A. Equations of motion for probability amplitudes}
We can find the time evolution of the probability amplitudes for the generic problem of multiple modes with three qubits trapped inside the optical cavity and up to three excitations in the system using the time-dependent Schr\"odinger equation $i\hbar\partial_t\ket{\Psi(t)}=\hat{\mathscr{H}}_{int}\ket{\Psi(t)}$. After performing a transformation into a rotating frame with the qubit transition frequency $\omega_{eg}$, we obtain the following set of coupled differential equations obeyed by the probability amplitudes (which we used for our numerical simulations):
\begin{widetext}
\begin{subequations}
\begin{align}
\frac{d\mathcal{A}_{123}(t)}{dt} = & (-3\gamma/2) \mathcal{A}_{123}(t)+\sum_i\sum_{j>i}\sum_{k\neq i,j}\sum_{\alpha} \mathscr{G}_{\alpha k}\mathcal{A}_{ij\alpha}(t),
\\
\left(\frac{d}{dt}+i\tilde\Delta_\alpha+\gamma\right)\mathcal{A}_{ij\alpha}(t)= & ~\sqrt{2}\mathscr{G}_{\alpha i}\mathcal{A}_{j\alpha\alpha}(t) + \sqrt{2}\mathscr{G}_{\alpha j}\mathcal{A}_{i\alpha\alpha}(t)- \mathscr{G}^*_{\alpha {k\neq i,j}} \mathcal{A}_{123}(t) + \sum _{\beta>\alpha} \mathscr{G}_{\beta_i}\mathcal{A}_{j\alpha\beta}(t) \nonumber\\
& + \sum _{\beta<\alpha} \mathscr{G}_{\beta i}\mathcal{A}_{j\beta\alpha}(t) + \sum _{\beta>\alpha} \mathscr{G}_{\beta j}\mathcal{A}_{i\alpha\beta}(t) + \sum _{\beta<\alpha} \mathscr{G}_{\beta j}\mathcal{A}_{i\beta\alpha}(t),\\
\left(\frac{d}{dt}+2i\tilde\Delta_\alpha+\gamma/2\right)\mathcal{A}_{i\alpha\alpha}(t)= & ~\sqrt{3}\mathscr{G}_{\alpha i}\mathcal{A}_{\alpha\alpha\alpha}(t) + \sum_{\beta>\alpha}\mathscr{G}_{\beta i}\mathcal{A}_{\alpha\alpha\beta}(t)+ \sum_{\beta<\alpha}\mathscr{G}_{\beta i}\mathcal{A}_{\beta\alpha\alpha}(t) -\sqrt{2}\sum_{j> i}\mathscr{G}^*_{\alpha j}\mathcal{A}_{ij\alpha}(t)\nonumber\\
&-\sqrt{2}\sum_{j< i}\mathscr{G}^*_{\alpha j}\mathcal{A}_{ji\alpha}(t),\\
\left(\frac{d}{dt}+i\tilde\Delta_\alpha+i\tilde\Delta_\beta\right)\mathcal{A}_{i\alpha\beta}(t) &= \sqrt{2}\mathscr{G}_{\alpha i}\mathcal{A}_{\alpha\alpha\beta}(t) + \sqrt{2}\mathscr{G}_{\beta i}\mathcal{A}_{\alpha\beta\beta}(t) + \sum_{\alpha<\beta<\gamma} \mathscr{G}_{\gamma i} \mathcal{A}_{\alpha\beta\gamma}(t) +\sum_{\alpha<\gamma<\beta} \mathscr{G}_{\gamma i} \mathcal{A}_{\alpha\gamma\beta}(t)\nonumber\nonumber\\
& + \sum_{\gamma<\alpha<\beta<} \mathscr{G}_{\gamma i} \mathcal{A}_{\gamma\alpha\beta}(t) - \sum_{j>i}\mathscr{G}^*_{\beta j}\mathcal{A}_{ij\alpha}(t)- \sum_{j<i}\mathscr{G}^*_{\beta j}\mathcal{A}_{ji\alpha}(t)-\sum_{j>i}\mathscr{G}^*_{\alpha j}\mathcal{A}_{ij\beta}(t)\nonumber\\
&-
\sum_{j<i}\mathscr{G}^*_{\alpha j}\mathcal{A}_{ji\beta}(t),\\
\left(\frac{d}{dt}+3i\tilde\Delta_\alpha\right)\mathcal{A}_{\alpha\alpha\alpha}(t)
= &- \sqrt{3}\sum_{i=1}^3 \mathscr{G}^*_{\alpha i} \mathcal{A}_{i\alpha\alpha}(t),\\
\left(\frac{d}{dt}+2i\tilde\Delta_\alpha+i\tilde\Delta_\beta\right)\mathcal{A}_{\alpha\alpha\beta}(t)
= &-\sum^3_{i=1}\mathscr{G}^*_{\beta i}\mathcal{A}_{i\alpha\alpha}(t)-\sqrt{2}\sum^3_{i=1}\mathscr{G}^*_{\alpha i}\mathcal{A}_{i\alpha\beta}(t),\\
\left(\frac{d}{dt}+i\tilde\Delta_\alpha+2i\tilde\Delta_\beta\right)\mathcal{A}_{\alpha\beta\beta}(t)
= &-\sum^3_{i=1}\mathscr{G}^*_{\alpha i}\mathcal{A}_{i\beta\beta}(t)-\sqrt{2}\sum^3_{i=1}\mathscr{G}^*_{\beta i}\mathcal{A}_{i\alpha\beta}(t),\\
\left(\frac{d}{dt}+i\tilde\Delta_\alpha+i\tilde\Delta_\beta+i\tilde\Delta_\gamma\right)\mathcal{A}_{\alpha\beta\gamma}(t)
= &- \sum^3_{i=1} \mathscr{G}_{\gamma i}^* \mathcal{A}_{i\alpha\beta}(t)- \sum^3_{i=1} \mathscr{G}_{\beta i}^* \mathcal{A}_{i\alpha\gamma}(t)- \sum^3_{i=1} \mathscr{G}_{\alpha i}^* \mathcal{A}_{i\beta\gamma}(t).
\end{align}
\end{subequations}
\end{widetext}

\bibliographystyle{ieeetr}
\bibliography{paper}
\end{document}